\newcommand{\lblroot}{dbcs}
\newcommand{\imgsroot}{images}

\newcommand{\rqOneDbcs}{Does the crowd provide enough support to calculate a trustworthy compatibility score?}
\newcommand{\rqTwoDbcs}{Which other sources of information can be considered when the crowd does not provide enough support to calculate a compatibility score?}
\newcommand{\rqThreeDbcs}{How much confidence should client packages place in the compatibility score?}

\newcommand{\rqone}{\rqOneDbcs}
\newcommand{\rqtwo}{\rqTwoDbcs}
\newcommand{\rqthree}{\rqThreeDbcs}

\newcommand{\rangecompatibilityscores}{origin version range compatibility scores\xspace}
\newcommand{\RangeCompatibilityScores}{Origin Version Range Compatibility Scores\xspace}
\newcommand{\clienthistorymetrics}{client history of updates\xspace}
\newcommand{\ClientHistoryMetrics}{Client History of Updates\xspace}

\newcommand{\code}[1]{\texttt{#1}}
\newcommand{\quotes}[1]{``#1''}
\newcommand{\fnu}[1]{\footnote{\url{#1}}}

\newcommand{\gh}{\textsf{GitHub}\xspace}

\newcommand{\pr}{PR\xspace}
\newcommand{\prs}{PRs\xspace}
\newcommand{\semver}{SemVer\xspace}
\newcommand{\db}{\textsf{Dependabot}\xspace}
\newcommand{\dbpr}{\db PR\xspace}
\newcommand{\dbprs}{\db \prs}

\newcommand{\checks}{\textsf{Checks}\xspace}
\newcommand{\travis}{\textsf{Travis CI}\xspace}
\newcommand{\ghactions}{\textsf{\gh Actions}\xspace}
\newcommand{\husky}{\textsf{Husky}\xspace}

\documentclass[10pt,journal,compsoc]{IEEEtran}

\usepackage[utf8]{inputenc} 
\usepackage[T1]{fontenc}
\usepackage{graphicx}
\usepackage{color} 
\usepackage[bookmarksdepth=2]{hyperref} 
\usepackage[misc,geometry]{ifsym}  
\usepackage{float} 
\usepackage[sort&compress,square,comma,numbers]{natbib}
\usepackage{amsmath,amssymb,amsfonts} 
\usepackage[table,xcdraw]{xcolor} 
\usepackage{multirow} 
\usepackage{mdframed} 
\usepackage{xspace} 
\usepackage{soul} 
\usepackage{subcaption} 
\usepackage{enumitem} 
\usepackage[many]{tcolorbox} 
\usepackage{rotating} 
\usepackage{tabularx} 
\usepackage{spverbatim} 
\usepackage{siunitx} 
\usepackage{booktabs} 
\usepackage{listings} 
\usepackage{textcomp} 
\usepackage{makecell}
\usepackage{titlecaps}

\newtcolorbox{mybox}[2][]{
    top=0.15in,left=4pt,right=4pt,bottom=4pt,
    fonttitle=\bfseries,
    colbacktitle=gray,
    colback=gray!5,
    colframe=gray!40!black,
    enhanced,
    attach boxed title to top left={xshift=1.5em,yshift=-\tcboxedtitleheight/2},
    boxed title style={size=small},
    drop shadow={black!50!white},
    title=#2,#1
}

\newcommand{\countobservations}{
    \def \countobservations{1}
}
\newcounter{observation}
\newenvironment{observation}[1]{
    \refstepcounter{observation}
    \noindent \textbf{Observation~\theobservation) \textit{#1}}
}
\countobservations

\newcommand{\countimplications}{
    \def \countimplications{1}
}
\newcounter{implication}
\newenvironment{implication}[1]{
    \refstepcounter{implication}
    \noindent \textbf{Implication~\theimplication ) \textit{#1}}
}
\countimplications

\newboolean{showcomments}
\setboolean{showcomments}{true}
\ifthenelse{\boolean{showcomments}}{
    \newcommand{\nbc}[3]{
        {\colorbox{#3}{\bfseries\sffamily\scriptsize\textcolor{white}{#1}}}
        {\textcolor{#3}{\sf\small$\blacktriangleright$\textit{#2}$\blacktriangleleft$}}
    }
} {
    \newcommand{\nbc}[3]{}
}

\newcommand{\tableFontSize}{\footnotesize}

\begin{document}
    \title{Leveraging the Crowd for Dependency Management: An Empirical Study on the Dependabot Compatibility Score}
	\author{Benjamin~Rombaut,~\IEEEmembership{}
	Filipe~R.~Cogo,~\IEEEmembership{}
	Ahmed~E.~Hassan~\IEEEmembership{}

	\IEEEcompsocitemizethanks{
	\IEEEcompsocthanksitem Benjamin Rombaut and Ahmed E. Hassan are with the Software Analysis and Intelligence Lab (SAIL) in the School of Computing at Queen's University, Kingston, Ontario, Canada.\protect\\
	Email: benjamin.rombaut@queensu.ca, ahmed@cs.queensu.ca
	\IEEEcompsocthanksitem Filipe R. Cogo is with the Centre for Software Excellence (CSE) at Huawei, Canada.\protect\\
	E-mail: filipe.roseiro.cogo1@huawei.com
	}
	}
    \IEEEtitleabstractindextext{%
	\begin{abstract}
	Software is increasingly being built by client packages making use of third-party provider packages in the form of dependency relationships, which means client packages must face the essential and risky task of keeping their provider package dependencies up-to-date. \db, a popular dependency management tool, includes a compatibility score feature that helps client packages assess the risk of accepting a dependency update by leveraging knowledge from \quotes{the crowd}. For each dependency update, \db calculates this compatibility score by dividing the number of successful updates by the total number of update attempts (candidate updates) made by client packages that use the provider package as a dependency. In this paper, our objective is to study the efficacy of leveraging the crowd to help client packages assess the risks involved with accepting a dependency update. To accomplish this, we analyze 579,206 pull requests opened by \db to update a dependency, along with 618,045 compatibility score records calculated by \db. 
	We find that the majority of compatibility scores do not have the minimum number of required candidate updates for the compatibility score badge to be shown on \db pull requests. When the compatibility scores do have enough candidate updates, the vast majority of the scores are above 90\%, suggesting that client packages should have additional angles to evaluate the risk of an update and the trustworthiness of the compatibility score. To overcome the lack of candidate updates when calculating a compatibility score, we propose metrics that amplify the input from the crowd and demonstrate the ability of those metrics to predict the acceptance of an update by client packages. We also verify that historical update metrics from client packages can be used to provide a more personalized compatibility score.
	Finally, we find that client packages should be hesitant to place total confidence in compatibility scores, as the candidate updates that are used to calculate the scores can be low both in quantity and quality. 
	Based on our findings, we argue that, when leveraging the crowd, dependency management bots should
	(i) be mindful of ways to amplify input from the crowd,
	(ii) consider historical metrics from the client package to provide a personalized compatibility score,
	(iii) include a confidence interval to help calibrate the trust clients should place in the compatibility score, and
	(iv) take into consideration the quality of tests that exercise candidate updates so as to avoid biasing the compatibility score.
	\end{abstract}

	}

	\maketitle
	\IEEEdisplaynontitleabstractindextext
	\IEEEpeerreviewmaketitle

	\section{Introduction}
	\label{\lblroot:intro} 
	Software is increasingly being built by making use of dependency relationships, where a client package relies on specific versions of a provider package. These provider packages enable code reuse and have been shown to improve developer productivity, software quality, and time-to-market of software products~\citep{abdalkareem_why_2017, abdalkareem_impact_2020}. However, clients must also incur the cost of managing these dependencies, as provider packages continuously release new versions containing bug fixes, new functionalities, and security enhancements~\citep{mirhosseini_can_2017, decan_impact_2018}.

	An important development decision that is faced by clients is whether to update the provider package from the presently used version in their package (i.e., the \textit{origin version}) to the newest release of the provider (i.e., the \textit{target version}). Doing so allows clients to receive the aforementioned potential benefits, but at the risk of these new versions modifying existing functionality or introducing API-backwards incompatibilities (a.k.a., breaking updates)~\citep{bogart_how_2016}. One strategy employed by client packages to protect against breaking updates is to run their own continuous integration (CI) pipeline, including unit and integration tests, against newly released versions of their dependencies~\citep{hilton_usage_2016}. Unless an update is intentionally breaking backwards compatibility (e.g., a major release), the client's CI pipeline should continue to pass with the new release applied~\citep{raemaekers_semantic_2017}. 

	However, many client packages do not have a full CI pipeline enabled~\citep{hilton_usage_2016}, and therefore are unable to automatically test whether a dependency update will be compatible with their package. One strategy that attempts to address this issue is to leverage knowledge from the crowd to provide insights about the risk of a newly released version of a provider package. In fact, \citet{mujahid_using_2020} and \citet{mezzetti_type_2018} both propose techniques that leverage the test suites of clients of a provider package in an effort to detect breaking changes in new releases of the provider package, and use these test outcomes as crowd-sourced indicators of the risk of adopting said provider release.

	\db\fnu{https://dependabot.com/} is an automated dependency management tool that packages on \gh\fnu{https://github.com/} can integrate with to automate the process of updating and testing new releases from provider packages. \db sits between a client’s package manager and \gh, observing all of the provider packages the client package depends on. Each time one of these providers release a new version, \db opens a pull request (\pr) in the client package with the dependency update, and the client's CI pipeline is run automatically on the updated dependency to test if it is a breaking update. \db records the result of updating the dependency, and calculates a \emph{compatibility score} for the provider package release as the percentage of \prs with a successful CI conclusion (\textit{successful updates}) to the total number of \prs updating between the origin and target versions of said provider (\textit{candidate updates}). This compatibility score is shown as a badge on \prs that are opened by \db for the same provider update, and is meant to give practitioners a sense of the involved risk when updating a dependency by leveraging the knowledge of \quotes{the crowd}, so that clients can be confident a new provider version is backwards compatible and bug-free. 

	However, this technique has its limitations, as it requires a crowd of a large scale to work effectively, and there is a lack of research that examines the value and challenges of using this approach. One the one hand, since \db is state-of-the-art and the most widely used automated dependency management bot leveraging the idea of crowd-based risk assessment, clients stand to reap the benefits of these indicator metrics to help them keep their dependencies up-to-date and their packages in working order. On the other hand, it is unknown whether the crowd is actually able to provide a strong enough signal for \db to be able to calculate trustworthy compatibility scores, nor the level of confidence clients should actually place in these compatibility scores. 

	Therefore, in this paper, we study the efficacy of \db's strategy of leveraging the crowd to provide a compatibility score to help clients assess the risks involved with dependency management. In the following, we list our research questions and key observations:

	\smallskip

	\noindent \textbf{RQ1) \rqone} We examine the proportion of compatibility scores that have the minimum number of candidate updates required by \db and the range of scores practitioners most often see when they receive a \dbpr.
	Our results indicate that compatibility scores tend to have a small number of candidate updates and are heavily skewed towards 100\%. Therefore, clients should be hesitant to trust compatibility scores, and other sources of information should be considered to calculate compatibility scores to overcome the lack of candidate updates.

	\smallskip

	\noindent \textbf{RQ2) \rqtwo} We examine seven features across two dimensions: i) \textit{\rangecompatibilityscores}, which considers the candidate updates from a range of origin versions of a provider package (e.g., \code{2.0.x}) that have been updated to a specific target version (e.g., \code{2.0.4}) and aims to amplify the knowledge from the crowd, and ii) \textit{\clienthistorymetrics}, which aim to capture the historical stability of the client’s package in general, the historical compatibility of the provider package with the client's package, and the historical level of confidence the client package places in the provider package. We observe that the range compatibility score dimension can help to increase the number of candidate updates that are used to calculate compatibility scores. We also find that features from both of the aforementioned dimensions can result in models that predict whether a dependency update will be accepted or rejected by a client package with an AUC of 0.64-0.80, with historical metrics from the client package tending to have the highest predictive power.

	\smallskip

	\noindent \textbf{RQ3) \rqthree} We evaluate the confidence \db has in compatibility scores by building an associated confidence interval for each compatibility score. We observe that half of compatibility scores with at least 5 candidate updates have a confidence interval whose bounds are further than 15\% from the compatibility score. 
	We also explore the quality of checks that make up the CI pipelines of candidate updates, and find that candidate updates that contribute to compatibility scores may not always truly test the associated dependency update.

	\smallskip

	The aforementioned results led us to conclude that, while popular dependency management bots like \db making use of the crowd to assess the compatibility of a dependency update is a promising strategy, the compatibility scores are often not available, and, even when the scores are available, can be misleading for clients without the support of a confidence interval. Additionally, bots should employ further methods to help amplify input from the crowd or consider historical upgrade metrics to assess whether a client package should accept or reject a dependency update.

	More generally, the main contributions of this paper are: 
	(i) an empirical study that examines \db's current strategy of leveraging the crowd to provide a compatibility score to help clients assess the risks involved with accepting a dependency update, 
	(ii) a description and evaluation of additional data sources that can be considered when the crowd does not provide enough support to calculate a compatibility score, 
	(iii) a description and evaluation of an approach to help calibrate the level of trust clients should place in the score, and
	(iv) a series of practical recommendations for designers of automated dependency management bots on effectively leveraging the crowd to help clients assess the risk of accepting a dependency update.

	The remainder of this paper is organized as follows. Section \ref{\lblroot:background} introduces key concepts related to our study. Section \ref{\lblroot:data} explains the employed data collection procedures. Section \ref{\lblroot:findings} presents the motivation, approach, and findings of our three research questions. Section \ref{\lblroot:discussion} discusses the implications of our findings. Section \ref{\lblroot:related_work} presents related work. Section \ref{\lblroot:threats} discusses the threats to the validity of our study. Finally, Section \ref{\lblroot:conclusion} concludes the paper.

	\clearpage

    \section{Background and Motivating Example}
	\label{\lblroot:background}
	In this section, we present the key concepts related to automated dependency management with \db (Section~\ref{\lblroot:background:dependabot}). We also present a motivating example (Section~\ref{\lblroot:background:motivating_example}) to help illustrate the intentions of the compatibility score.

	\subsection{Dependabot}
	\label{\lblroot:background:dependabot}
	More and more packages are making use of tools that help to automate dependency management. For example, bots are increasingly being used to notify client packages in the form of \prs when one of their dependencies releases a new version. Clients can then configure a CI pipeline to automatically test the new dependency release in an attempt to verify that it is compatible with their package and does not contain any API-backwards incompatibility or introduce other regressions.

	\db is perhaps currently the most popular automated dependency-management tool, having first launched on May 26, 2017\fnu{https://dependabot.com/blog/introducing-dependabot/} and later being acquired by \gh on May 23, 2019\fnu{https://dependabot.com/blog/hello-github/}. \db supports a wide range of different language ecosystems, including JavaScript, Ruby, and Python to name a few. \db sits between a package manager and \gh, observing all of the providers a client package depends on. Each time one of these providers release a new version, \db opens a new \pr with the client's dependency specifications updated to accept the newly released provider version. Once a \dbpr is created, the client's CI pipeline, if configured, runs automatically against the \pr branch to determine if the new version of the provider passes all of the client's tests. The client can then decide whether they would like to accept or reject the \dbpr.

\begin{figure}[!htb]
		\centering
		\includegraphics[width=\linewidth]{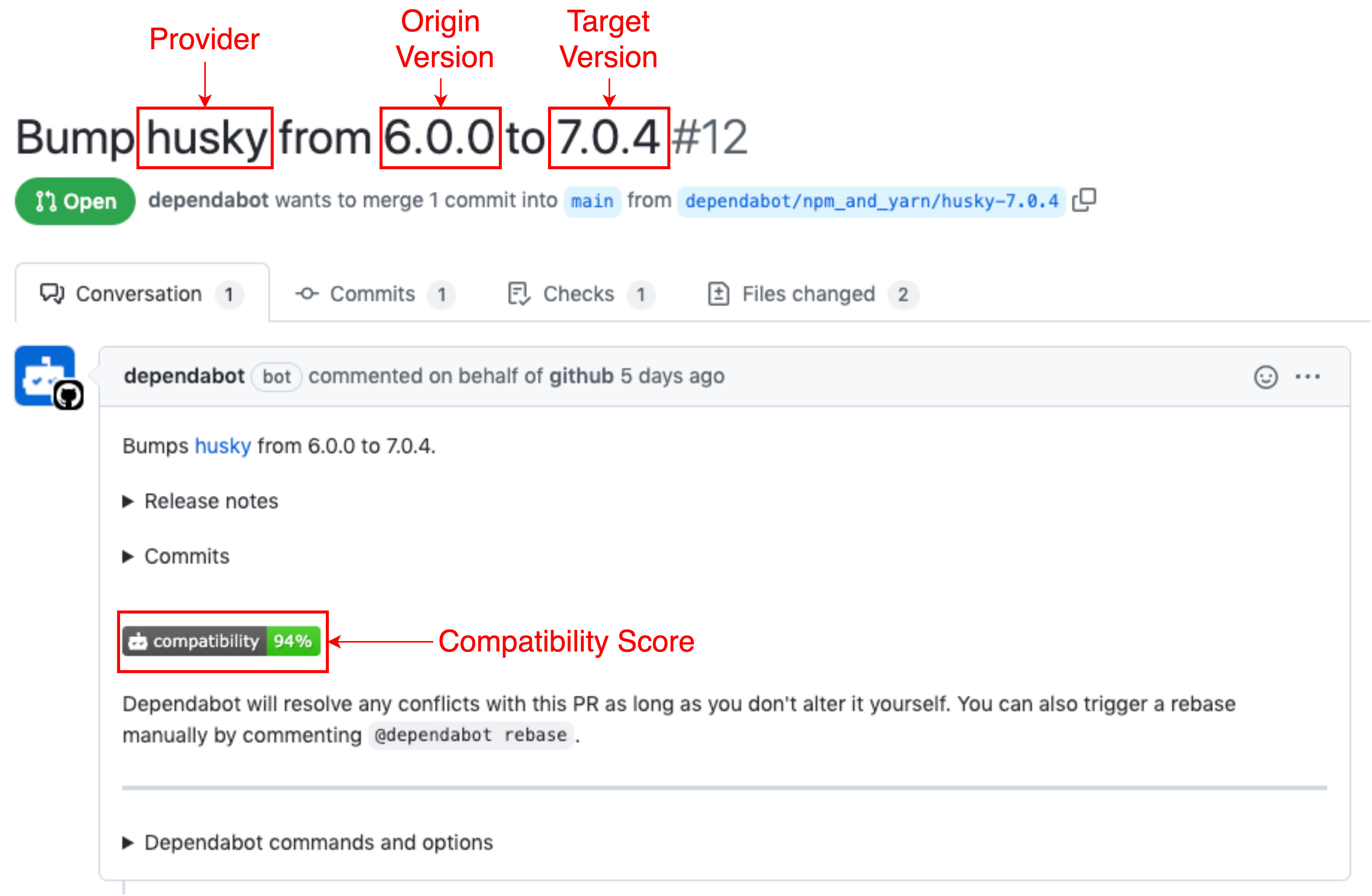}
		\caption{
			An example of a \dbpr with the provider package name, origin version, target version, and the compatibility score highlighted.}
		\label{\lblroot:fig:dbpr_example}
	\end{figure}

	To support clients in their decision of whether they should accept or reject a \dbpr, \db includes a summary statistic called a \emph{compatibility score} that leverages knowledge from the crowd to provide insights about the risk of a newly released version of a provider package. When a new provider version is released, \db creates similar \prs across multiple client packages to update the provider from the origin version used by each client to the target version which has been newly released by the provider. More formally, the provider package named $P$, origin version $V_O$, and target version $V_T$ create a 3-tuple for the dependency update ($P$, $V_O$, $V_T$). For each client with CI enabled (e.g., \travis\fnu{https://travis-ci.com/} or \ghactions\fnu{https://docs.github.com/en/actions}) and a previously passing test suite, \db records whether the 3-tuple dependency update breaks any of the client's tests. \db considers \prs that meet this criteria to be \emph{candidate updates}. \db considers a candidate update to be a \emph{successful update} if the client's CI pipeline is in a passing state with the dependency update. Figure~\ref{\lblroot:fig:dbpr_example} provides an example of a \dbpr with the provider package name, origin version, target version, and the compatibility score highlighted.

	The compatibility score of a dependency update is the percentage of CI runs that passed when updating the dependency between the same origin and target versions (i.e., the number of successful updates divided by the number of candidate updates for the origin version and target version of the provider package). It is important to note that the client does not necessarily have to merge the \dbpr in order for it to be considered a candidate update and contribute to the compatibility score. In order for the compatibility score to correctly show up on the \dbpr, the dependency update must have at least 5 candidate updates\fnu{https://github.com/dependabot/dependabot-core/issues/4001\#issuecomment-870399478}. Otherwise, the badge will simply say that the compatibility score is \quotes{unknown}.

	\subsection{Motivating Example}
	\label{\lblroot:background:motivating_example}
	To help illustrate how the compatibility score is used in practice and how it can be misleading in the context of dependency management, we provide a simple motivating example.

	Alice is a software developer responsible for developing and maintaining an application in her company. In order to enable code reuse and speed up development time, she relies on a few third-party packages to accomplish specific tasks in her application. However, in order to keep up with the demanding timeline of her employer, she has only managed to build a CI testing pipeline that amounts to \quotes{smoke tests}\fnu{https://softwaretestingfundamentals.com/smoke-testing/} (i.e., a non-exhaustive set of tests that aim at ensuring only the most important functions of her application work). She has not written any tests that exercise the portions of her application that make use of her dependencies, as she figures that these packages would be \quotes{deployment tested} (i.e., they are extensively used in production by many clients).

	As with many provider packages, the dependencies Alice uses get updated frequently. Alice wants to be more proactive in managing her software dependencies, so she uses \db to automatically open \prs to update her dependencies as new releases become available. She knows that \db is the most commonly used automated dependency management bot and enjoys the convenience of being notified when her dependencies become out-of-date. 

	While Alice very much wants to keep her dependencies up-to-date, she is aware of the risks involved with blindly accepting a new update. She has heard stories from other developers who have had to drop all of their work in order to fix a broken CI pipeline caused by a dependency update. Even worse, Alice has even heard of developers who were only informed by their customers that their application was broken weeks after deploying a new version that contained a breaking dependency update. She could only imagine the amount of work that it took to find that this particular dependency update was the root cause, not to mention the user's perceived lack of quality that comes with deploying a broken version of the application. Since Alice knows her test suite does not sufficiently cover her application, she thinks the compatibility score badge \db includes on the \prs is a very helpful indicator for the compatibility of the dependency update, and tends to rely on it when deciding whether or not to accept a dependency update.

	One day, Alice sees that \db has opened a \pr in her application. After briefly examining the \pr, she sees that her CI tests pass when applied against the dependency update, but that \db has not been able to calculate a compatibility score for the update. Knowing that her tests are most likely not capable of exercising major portions of her dependencies, she decides to hold off on taking any action on this \pr. 

	After a few days, Alice checks back on the \pr, and finds that \db reports that updating the dependency from the version that she currently uses to the newly released target version has a compatibility score of 100\%. With this information in mind, she decides to merge the \pr.

	A few days later, Alice gets a message from her boss stating that their application is experiencing some unexpected behaviour. After debugging, Alice finds that the recent change she made by merging the \dbpr for the dependency update introduced the issue. Even though her CI testing pipeline passed, the tests were not able to detect the breaking behaviour in the updated dependency - the tests simply did not cover the case causing the unexpected behaviour. Remembering the 100\% compatibility score she saw when she merged the \pr, she investigates and discovers that the dependency update only had 5 candidate updates - one of which was actually the \pr \db opened for her own application! Alice's confidence in the compatibility score has been severely shaken after this incident. She made the wrong decision because she didn't know \quotes{how much} she could trust the compatibility score when she merged the \dbpr, and now no longer believes the compatibility score to be a reliable metric.

	\section{Data Collection}
	\label{\lblroot:data}
\begin{figure}[!htb]
		\centering
		\includegraphics[width=0.94\linewidth]{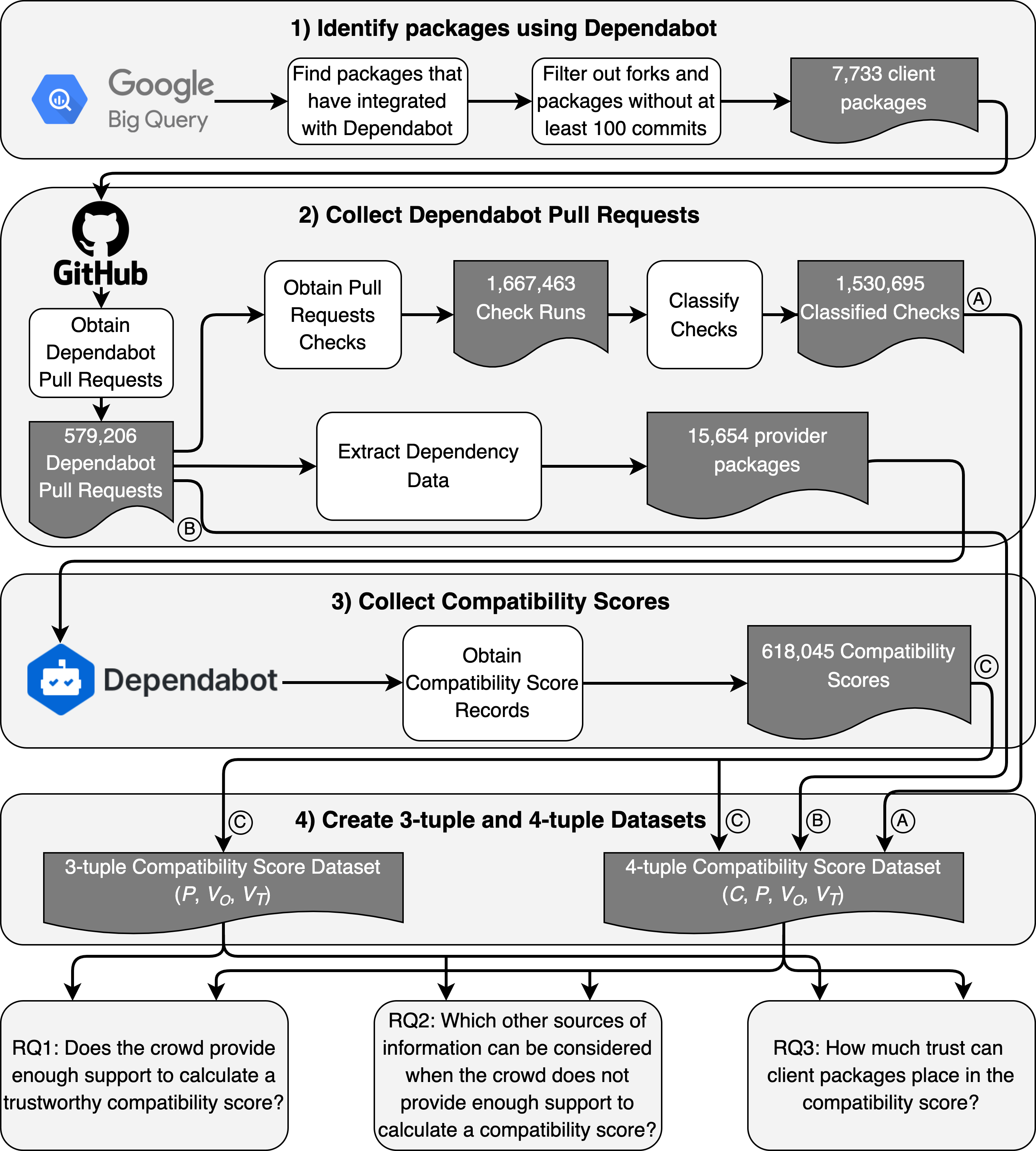}
		\caption{
			Overview of the data collection process.}
		\label{\lblroot:fig:data_collection_diagram}
	\end{figure}

	In this section, we discuss how we collect the dataset to address the RQs outlined in the introduction. We use the workflow of Figure~\ref{\lblroot:fig:data_collection_diagram}: (i) we identify packages on \gh that use \db, (ii) we collect all \dbprs for each packages that is identified in the previous step, and extract the necessary information and collect related artifacts for each \dbpr, (iii) we collect the compatibility scores for the provider package updates that are related to each \dbpr identified in the previous step, (iv) we build two distinct datasets using the data collected in the two previous steps. 

	Next, we provide a more in-depth explanation of each step in our data-collection workflow.

	\subsection{Identify packages using \db}
	\label{\lblroot:data:project_selection}
	To identify the packages using \db, we leverage the \textsf{Google BigQuery Public Datasets}\fnu{https://cloud.google.com/bigquery/public-data/} to search for commits on \gh that have been authored by \db. Each of these commit records contains the parent package name on \gh, which we use to build our list of packages to include in our study. It is known that \gh contains some toy packages~\citep{kalliamvakou_promises_2014} which are not representative of the software packages we aim to investigate. Therefore, once the dataset of packages using \db is collected, we apply some filtering criteria for selecting a set of packages with a history of activity. We only include packages that are non-forked and contain at least 100 commits, as recommended by prior studies~\cite{alfadel_use_2021, mirhosseini_can_2017, kalliamvakou_promises_2014}. In total, we extract a list of 7,733 \gh packages that meet our filtering criteria. Due to our filtering criteria, this is by no means an exhaustive list of all packages that use \db.

	\subsection{Collect Dependabot Pull Requests}
	\label{\lblroot:data:dependabot_pull_requests}
	We use the \gh API\fnu{https://docs.github.com/en/rest} to retrieve all \dbprs opened in the list of packages that we collected in Section~\ref{\lblroot:data:project_selection}. This step is necessary as a follow up to Section~\ref{\lblroot:data:project_selection} as it allow us to extract the information about the provider \db is attempting to update with the \pr. Overall, we collect a total of 579,206 \prs opened by \db for the time period between June 2017 and June 2021.

	\db includes information on the updated provider package in the title of the \pr (see Figure~\ref{\lblroot:fig:dbpr_example}). We extract the provider package name, the origin version of the provider used by the client, and the newly released target version of the provider from the \pr title using a set of regular expressions, which we use to collect compatibility scores in Section~\ref{\lblroot:data:compatibility_scores}. We are able to extract this information from 575,860 (99.4\%) of the \prs. Upon closer examination of the \prs for which we are not able to extract this information for, we find that these \prs are not in fact dependency updates, but rather automatic \prs created by \db to create or modify the \db configuration file in the client package, or to update \db itself.

	To determine whether a client's CI pipeline passed or failed when testing each dependency update, we retrieve the \gh \checks\fnu{https://docs.github.com/en/rest/reference/checks} that ran on each \pr. \checks are pipeline runs or custom scripts that perform specific tasks (e.g., linters\fnu{https://github.com/collections/clean-code-linters} or \travis builds), and they are used by \db to determine the result of a candidate update (i.e., success or failure). There can be multiple checks that run against a \pr. For example, a client might have a check that ensures the client's package builds, another check to run the test suite, and a final check to detect and fix any code style issues. The set of checks that run against a \pr make up the CI pipeline for the client's package. Overall, we find that 38\% of \dbprs (or 43\% of client packages) have a configured CI pipeline (i.e., a set of checks) to run on new \prs, which is in line with prior results by \citet{hilton_usage_2016} when examining the percentage of open-source packages that have a CI pipeline configured.

	\citet{hejderup_can_2021} find in their study that it is common for clients to have a low quality set of tests that run against dependency updates. With these findings in mind, we decide to classify the types of checks to determine what types of CI pipelines are run against \dbprs. We use the name of the check, which is used to give a high-level description of the task the check performs, to assign each check to a specific overarching category. We match 91.8\% of the checks using the process described in Appendix~\ref{\lblroot:appendix:ci_name_regex} to one of the following six categories: \textit{Build} (58.1\%), \textit{Test} (17.2\%), \textit{Useless} (11.2\%), \textit{Lint} (7.0\%), \textit{Deploy} (4.9\%), and \textit{Security Analysis} (1.6\%). We find that clients typically group their entire build, test, and deploy pipeline into a single check workflow, which explains why the \textit{Build} category is the most common. The \textit{Useless} category consists of checks that do not help with determining whether the changes contained in the \pr are compatible with the client package (e.g., automatically adding a label to the \pr or uploading build logs to a separate repository).

	\subsection{Collect Compatibility Scores}
	\label{\lblroot:data:compatibility_scores}
	For each specific provider package we extracted in Section~\ref{\lblroot:data:dependabot_pull_requests}, we retrieve all compatibility scores using the \db API\fnu{https://dependabot.com/compatibility-score/}. The \db API requires a package manager and a provider package name as query parameters, and returns the compatibility score records for all 3-tuple update combinations that \db has recorded for that provider package. Each compatibility score record contains the number of candidate updates and the number of successful updates \db has recorded for the 3-tuple dependency update in question. Overall, we collect the compatibility score records for a total of 618,045 3-tuple dependency updates.

	\subsection{Create 3-tuple and 4-tuple Datasets}
	\label{\lblroot:data:subsec:datasets}

	Because our package list in Section~\ref{\lblroot:data:project_selection} is a non-exhaustive set of package that use \db, the \dbprs we collect in Section~\ref{\lblroot:data:dependabot_pull_requests} do not represent the full list of the number of candidate updates that are used by \db to calculate the compatibility scores. Therefore, the process described in Section~\ref{\lblroot:data:compatibility_scores} is necessary to get the complete picture in terms of the number of candidate and successful updates contributing to the associated compatibility score for each 3-tuple update. Hence, we refer to the compatibility scores we collect in Section~\ref{\lblroot:data:compatibility_scores} as the \quotes{3-tuple dataset}.

	However, records from the 3-tuple dataset only contain the compatibility scores for the 3-tuple update in question, without any relating information on the specific \dbprs opened in client packages that are contributing as candidate or successful updates to these scores. As a result, we are not able study the relationship between compatibility scores and the associated merge status of candidate update \dbprs using the 3-tuple dataset.

	Therefore, we link the compatibility score from the 3-tuple dataset with the associated \dbprs (if present) we collected in Section~\ref{\lblroot:data:dependabot_pull_requests}. This allows us to link specific candidate update \dbprs to a compatibility score and provides a means to study the relationship between the compatibility scores and the merge status of said \dbprs. With the specific client linked to the 3-tuple dependency update for the compatibility score, we form a 4-tuple consisting of ($C$, $P$, $V_O$, $V_T$), where $C$ is the client package from a \dbpr, $P$ is a provider package used by $C$, $V_O$ is the origin version of $P$ used by $C$ at the time the \dbpr was opened, and $V_T$ is the newly released target version of $P$ at the time the \dbpr was opened (notice that $P$, $V_O$, and $V_T$ form a single record from the 3-tuple dataset). Hence, we refer to the this dataset as the \quotes{4-tuple dataset}.

	An additional description on how and why the 3-tuple and 4-tuple datasets may differ is included in Appendix~\ref{\lblroot:appendix:tuple_datasets}.

	\section{Findings}
	\label{\lblroot:findings}
	In this section, we present the results for each of our RQs. For each RQ, we discuss the motivation, the approach we used to address the RQ, and our findings.

	\subsection{RQ1: \rqone}
	\label{\lblroot:findings:existing_score}
	\noindent \textbf{Motivation.}
	Client packages often want to be aware of the risk of a dependency update breaking the build~\citep{bogart_when_2015}, particularly when the quality of test suites cannot be fully trusted -- a relatively common scenario according to recent research~\citep{hejderup_can_2021}. Dependency bots (e.g., \db) have recently integrated a new feature that leverages crowd-sourced information to estimate the risk of an update in the form of a compatibility score. However, the viability of the compatibility score has yet to be studied in practice, and it is unclear whether \db does in fact create enough dependency updates to be able to effectively determine a consensus from the crowd about whether a dependency update is safe or not, as well as whether client packages can rely on this consensus. In fact, these issues have been the source of complaints on the \db repository\fnu{https://github.com/dependabot/dependabot-core/issues/2443} as well as developer blog sites\fnu{https://dev.to/lhuria94/comment/ofe5}. Therefore, in this research question, we look to answer 1) how often do compatibility scores have the minimum number of candidate updates required to be shown as a badge on \dbprs? and 2) when the badge is shown on \dbprs, is the distribution of scores seen by client packages useful to assess the risk of an update?

	\smallskip \noindent \textbf{Approach.}
	To determine how often a known compatibility score shows up on \dbprs, we examine the proportion of candidate updates each dependency update has. Recall that in order for a known compatibility score badge to show up on the \pr, the dependency update must have at least 5 candidate updates. Otherwise, the badge will simply say that the compatibility score is \quotes{unknown}. We then examine the distribution of compatibility scores with at least 5 candidate updates to determine the range of scores practitioners most often see when they receive a \dbpr. We perform this analysis on the compatibility score for both the 3-tuple and 4-tuple datasets.

	\smallskip \noindent \textbf{Findings.}
	\begin{observation}{The majority (83\%) of dependency updates do not have enough candidate updates to display a compatibility score badge on Dependabot \prs.}
	Figure~\ref{\lblroot:fig:candidate_updates_distributions} shows the distributions of candidate updates for compatibility scores with at least 1 candidate update from both the 3-tuple and 4-tuple datasets. When examining the distribution of candidate updates for compatibility scores from the 3-tuple dataset, we find that only 17\% have at least 5 candidate updates. This finding was surprising, as it shows that, even though \db may be opening hundreds of \prs when a provider package releases a new version, more than four-fifths of the associated compatibility scores are still not shown on these \prs simply because the \quotes{consensus from the crowd} doesn't exist for the dependency update. In reality, this proportion is likely lower, as all compatibility score records in our 3-tuple dataset must have at least 1 candidate update (see Section~\ref{\lblroot:data:compatibility_scores}), which means we do not include the compatibility scores for dependency updates that have no candidate updates in our analysis.

	We find that approximately two-fifths (43\%) of the compatibility scores from the 4-tuple dataset do not have enough candidate updates for the compatibility score to show up on \dbprs, with the median number of candidate updates being 41. Recall that compatibility scores from the 4-tuple dataset include compatibility scores for dependency updates from provider packages used by active clients, as well as commonly used origin and target versions of these provider packages. Therefore, we expect these compatibility scores to have a higher number of candidate updates than those from the 3-tuple dataset. Although we observe this improvement compared to the 3-tuple dataset, it must be considered that, while these may be popular dependency updates, only 57\% have a compatibility score with enough candidate updates to be shown on the associated \dbpr. Our observations suggest that alternative sources of information should be considered to support dependency updates when the crowd does not provide enough input to calculate a compatibility score.

	\begin{figure}[ht]
		\centering
		\includegraphics[width=\linewidth]{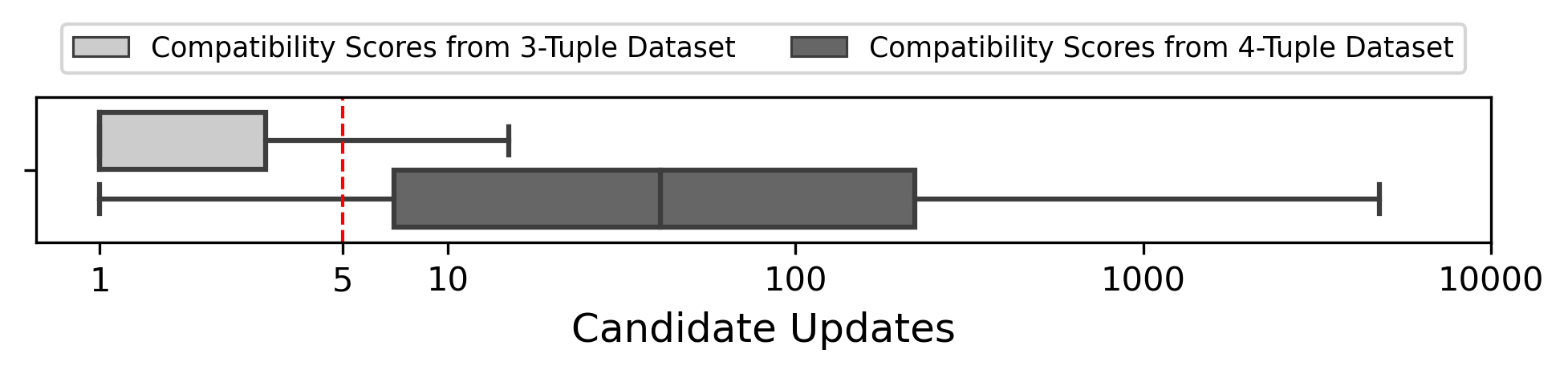}
		\caption{The distribution of candidate updates for compatibility scores with at least 1 candidate update from the 3-tuple and 4-tuple datasets.}
		\label{\lblroot:fig:candidate_updates_distributions}
	\end{figure}
	\end{observation}

	\smallskip \begin{observation}{Client packages are usually forced to distinguish between only a small range of compatibility scores.}
	When the badge with the compatibility score is shown on the \dbpr (i.e., the compatibility score has at least 5 candidate updates), we find that the vast majority of compatibility scores (76\% and 89\% in the 3-tuple and 4-tuple datasets, respectively) are greater than 90\%.

	Figure~\ref{\lblroot:fig:score_distributions} shows the distributions of compatibility scores that have at least 5 candidate updates from both the 3-tuple and 4-tuple datasets. We can see that, with so many compatibility scores grouped at the high end of the score range, it can be difficult for client packages to distinguish between such a small range of scores, and in fact may be mislead into thinking dependency updates are more compatible than they actually are. In order to help calibrate clients' trust on the usually excessively high compatibility scores, additional supporting metrics, such as the adoption of an accompanying confidence score for each compatibility score, might be useful.

	\begin{figure}[ht]
		\centering
		\includegraphics[width=\linewidth]{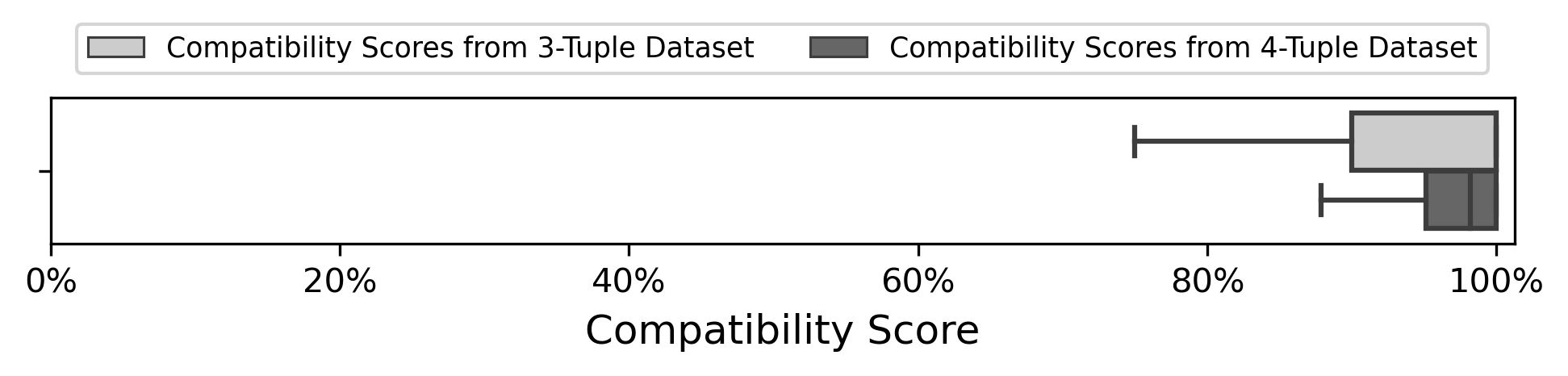}
		\caption{The distribution of compatibility scores that have at least 5 candidate updates from the 3-tuple and 4-tuple datasets.}
		\label{\lblroot:fig:score_distributions}
	\end{figure}
	\end{observation}

	\begin{mybox}{\parbox{8cm}{RQ1: \rqone}}
		\medskip
		\begin{itemize}[topsep = 6pt, itemsep = 3pt, label=\textbullet, wide = 0pt]
			\item The majority of compatibility scores do not have the minimum number of candidate updates to be shown correctly on \dbprs.
			\item The vast majority of the shown scores are above 90\%, hindering clients' ability to differentiate the risks of a dependency update.
		\end{itemize}
	\end{mybox}

	\subsection{RQ2: \rqtwo}
	\label{\lblroot:findings:other_metrics}
	\noindent \textbf{Motivation.}
	We found in RQ1 that the majority of dependency updates do not have enough candidate updates recorded by \db to correctly show a compatibility score badge on \prs. \db is the most popular dependency management bot available in the open-source community and it is unlikely that there is another tool that will be able to sample a crowd as large as the one available to \db, which presents a real issue with the concept of using the crowd to assess the risk of a dependency update. Therefore, \db should make use of metrics other than the number of candidate and successful updates for a specific origin and target version of a provider package when attempting to support clients in dependency management, especially when the number of candidate updates is low.

	\smallskip \noindent \textbf{Approach.} 
	Our goal is to explore alternative metrics that \db could take into account to provide a sense of the compatibility of a new dependency version in a client package, particularly when the crowd does not provide enough support to calculate a compatibility score for the dependency update. We examine 7 metrics divided into two dimensions (\textit{\rangecompatibilityscores} and \textit{\clienthistorymetrics}) that can be calculated using data already available to \db (summarized in Table~\ref{\lblroot:table:dimension_features}).

	\begin{table*}[ht]
	\tableFontSize
	\caption{
		Dimensions and their features that are used to assess the compatibility of a new dependency version in a client package when the crowd does not provide enough support to calculate a compatibility score for the dependency update.
	}
	\label{\lblroot:table:dimension_features} 
	\begin{center}
	\begin{tabularx}{\linewidth}{
		>{\hsize=.2\hsize}X
		>{\raggedright\hsize=.3\hsize}X
		>{\hsize=.4\hsize}X
		>{\hsize=.4\hsize}X
	}
	\toprule
	\textbf{Dimension} & \textbf{Metric Name} & \textbf{Rational} & \textbf{Description} \\
	\toprule
	\multirow{3}{10em}{\RangeCompatibilityScores} & 
		Patch Origin Version Range Compatibility Score & 
		Amplify input from the crowd by considering candidate updates from similar origin versions. & 
		The compatibility score for provider package $P$ calculated using the candidate updates from all 3-tuples matched in ($P$, \code{x.y.*}, \code{x.y.z}). \\
	& 
		Minor Origin Version Range Compatibility Score & 
		Amplify input from the crowd by considering candidate updates from origin versions which may be less similar than patch ranges, but still should not contain breaking changes. & 
		The compatibility score for provider package $P$ calculated using the candidate updates from all 3-tuples matched in ($P$, \code{x.*.*}, \code{x.y.z}). \\
	& 
		Major Origin Version Range Compatibility Score & 
		Maximize amplifying input from the crowd by incorporating candidate updates from all origin versions. & 
		The compatibility score for provider package $P$ calculated using the candidate updates from all 3-tuples matched in ($P$, \code{*.*.*}, \code{x.y.z}). \\ 
	\midrule
	\multirow{2}{10em}{\ClientHistoryMetrics} & 
		Passing \dbprs & 
		Captures the historical stability of the client's package in general. & 
		The number of \dbprs previously opened in the client package that have successfully passed the client's CI pipeline. \\
	& 
		Passing Provider \dbprs & 
		Captures the historical stability between the client package and the provider package \db is opening a \pr to update. & 
		The number of \dbprs for the same provider previously opened in the client's package that have successfully passed the client's CI pipeline. \\
	& 
		Merged \dbprs &
		Captures the level of trust the client package has with \dbprs in general. & 
		The number of \dbprs previously merged by a user in the client's package. \\ 
	& 
		Merged Provider \dbprs & 
		Captures the level of trust the client package has in the provider package \db is opening a \pr to update. & 
		The number of \dbprs for the same provider previously merged by a user in the client's package. \\ 
	\bottomrule
	\end{tabularx}
	\end{center}
	\end{table*}

	\textbf{\RangeCompatibilityScores:} Because \db exclusively counts candidate updates with the same origin and target version of a provider towards a compatibility score, the number of candidate updates for each compatibility score is severely limited. While there may be a high number of candidate updates for the provider package overall, all of these candidate updates end up being spread across a wide range of potential origin version and target version combinations.

	To address this issue, we consider using the candidate updates from a range of origin versions that have been updated to a specific target version. We take inspiration from the semantic versioning (\semver) scheme\fnu{https://semver.org}, a popular policy for communicating the type of changes made to a software package, where clients can specify whether they would like to accept a range of versions from the provider\fnu{https://nodesource.com/blog/semver-tilde-and-caret/}~\citep{dietrich_dependency_2019, decan_what_2020}. Similarly, we calculate three origin version range compatibility score metrics: a \textit{patch origin verion range compatibility score}, a \textit{minor origin version range compatibility score}, and a \textit{major origin version range compatibility score}. These origin version range compatibility scores are calculated in the same way as the raw compatibility score (i.e., the number of successful updates divided by the number of candidate updates), but they each consider an increasingly wider range of origin versions of the provider package to select candidate updates from. 
	The patch origin version range compatibility score considers all origin versions of a provider package where only the patch version number of the origin version may differ. 
	The minor origin version range compatibility score considers all origin versions of a provider package where the minor or patch version numbers of the origin version may differ. Finally, the major origin version range compatibility score considers all origin versions of a provider package where the major, minor or patch version numbers of the origin version may differ (i.e., all origin versions of a provider package that have been updated to a specific target version).
	Table~\ref{\lblroot:table:dimension_features} provides an example of these matching patterns. It can be seen that the major origin version range compatibility score will match more 3-tuple updates than the minor origin version range compatibility score, which in turn will match more 3-tuple updates than the patch origin version range compatibility score. These metrics aim to amplify the input from the crowd by expanding the range of considered candidate updates for each compatibility score at the cost of generalizing the exact origin version of the provider being considered for each origin version range compatibility score.

	\textbf{\ClientHistoryMetrics:} Even when considering the candidate updates from a range of origin versions, there still might not be enough support from the crowd to reliably calculate a compatibility score. Therefore, we turn to historical metrics from the client package to help assess the risk involved with a dependency update. Specifically, we look at the number of \dbprs previously opened that passed the CI pipeline in the client package (both overall and for each specific dependency). These metrics aim to capture the historical stability of the client's package in general, as well as the historical compatibility between the client package and the provider package that the \dbpr is attempting to update.

	Additionally, we consider the number of \dbprs that have previously been merged in the client package (both overall and for each specific dependency). These metrics aim to capture the level of trust the client has in the specific provider package \db is attempting to update and the client's overall providers in general, as a higher number of merged \dbprs suggests a higher level of trust by the client.

	To investigate how well the individual dimensions can assess the compatibility of a dependency (i.e., their predictive power), we built a random forest model for each of the previously discussed dimensions, setting the dependent variable as whether the \dbpr is merged by a client package developer. We select whether the client package developer accepts or rejects the dependency update rather than, for example, the result of the client's CI pipeline running against the dependency update, because it is common for CI pipelines to contain low-quality tests~\citep{hejderup_can_2021}, which would bias how we assess the compatibility of the dependency update (e.g., low-quality CI pipelines might fail often due to flaky tests). Using whether a client package developer decides to merge the \dbpr allows us to capture whether deliberate action was taken by a human to either accept or reject the dependency update, as client package developers may be aware of issues with their CI pipeline and merge \dbprs with failed CI pipelines anyway because they know their pipeline failed for reasons unrelated to the dependency update. In fact, we found this to be the case in 28\% of \dbprs with a failed CI pipeline, where the client decides to merge the \dbpr anyway.

	When building these models, we only consider \dbprs from the 4-tuple dataset that have fewer than 5 candidate updates, as these are the cases where we have recorded the \dbpr and the crowd has not provided enough support for \db to calculate a reliable compatibility score. As a baseline, we build a random forest model with the raw compatibility score as the sole independent variable and the dependent variable as the merge result of the \dbprs. For our baseline model, we only consider \dbprs from the 4-tuple dataset set that have at least 5 candidate updates. We use the \code{ranger}\fnu{https://cran.r-project.org/web/packages/ranger/ranger.pdf} package in \code{R} as our random forest implementation due to its enhanced performance.

	To validate the performance and stability of our built models, we performed 100 out-of-sample bootstrap iterations to compute the median AUC (Area Under the receiver operator characteristics Curve) for each model. Prior work~\citep{tantithamthavorn_empirical_2017, lee_empirical_2020} has shown that the out-of-sample bootstrap technique had the best balance between the bias and variance of estimates. The out-of-sample bootstrap technique randomly samples data with replacement for $n$ iterations. The sampled data in an iteration is used as the training set for that iteration, while the data that was not sampled in that iteration is used as the testing set for that iteration. We then trained a model with the training set and calculated the AUC of the model with the testing set for each iteration.

	In addition, to investigate how well both of the studied dimensions can help to assess the compatibility of a dependency, we built a random forest model using all 7 metrics from both dimensions previously discussed. We evaluated the performance of this combined model using the same aforementioned process of computing the median AUC of the model with 100 out-of-sample bootstrap iterations.

\begin{figure}[ht]
		\centering
		\includegraphics[width=\linewidth]{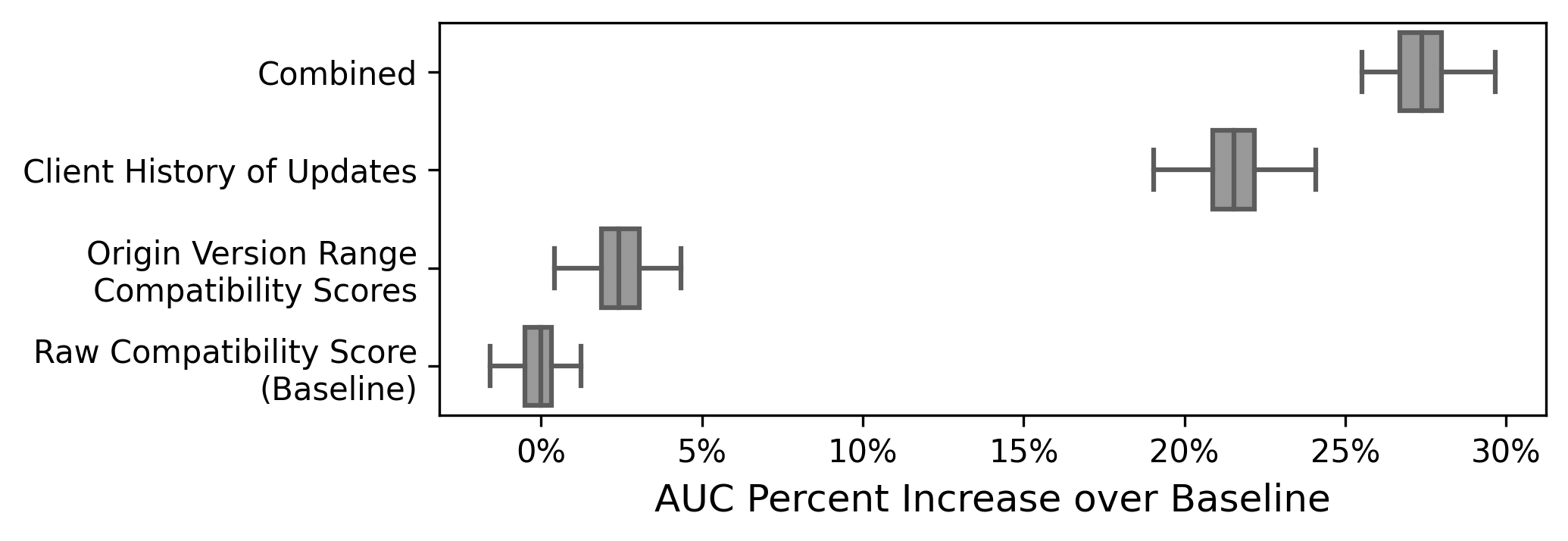}
		\caption{The distribution of the improvement in AUCs of models constructed with an individually studied dimension, and with all studied dimensions combined, compared against the baseline model.}
		\label{\lblroot:fig:model_auc_distributions}
	\end{figure}

	\smallskip \noindent \textbf{Findings.}
	\begin{observation}{There is room for improvement when establishing the relationship between the compatibility score for a dependency update and whether the associated Dependabot \pr is merged by the client package.}
	We observe that our baseline model built with the compatibility score as the sole predictor variable only achieves a median AUC of 0.62 (Figure~\ref{\lblroot:fig:model_auc_distributions} shows the distribution of AUC improvements compared to the median AUC for this baseline model). This shows that there is room for improvement when establishing the relationship between the compatibility score and the result of whether or not the client merged the \dbpr, and that it could be beneficial for \db to consider further metrics when trying to convey how compatible a dependency update really is for client packages.
	\end{observation}

	\begin{observation}{Considering a range of origin versions for a specific target version can help increase the number of candidate updates used to calculate the compatibility score.}
	While we find that the majority of compatibility scores from the 3-tuple dataset do not see any increase in the number of candidate updates used to calculate a compatibility scores when considering the patch origin version range (although the third quartile see 3x the number of candidate updates), the minor and major origin version range compatibility scores are able to consider respectively 5x and 10x the number of candidate updates as what is used by the raw compatibility score. We see relatively smaller improvements in the 4-tuple dataset, with the patch, minor, and major origin version range compatibility scores seeing respectively a median of 1x, 1.5x, and 1.9x the number of candidate updates as what is used by the raw compatibility score. Figure~\ref{\lblroot:fig:range_candidate_updates_by_dependency_update_type} shows the  distribution  of  ratios  of  candidate  updates  for each  origin version  range  compatibility  score  to  the  associated original  compatibility  score  from  the  3-tuple  and  4-tuple datasets.

	Considering a range of origin versions for a specific target version can also help to increase the number of compatibility scores that meet the required threshold number of candidate updates (i.e., 5) for \db to display the badge on the associated \pr.
	Recall from RQ1 that only 17\% of compatibility scores from the 3-tuple dataset have at least 5 candidate updates, while 83\% of compatibility scores from the 4-tuple dataset have at least 5 candidate updates. 
	When we consider our calculated origin version range compatibility scores for the 3-tuple dataset, we find that 39\%, 68\%, and 78\% of patch, minor, and major \rangecompatibilityscores respectively have at least 5 candidate updates. 
	For the 4-tuple dataset, we find that 86\%, 90\%, and 92\% of patch, minor, and major \rangecompatibilityscores respectively have at least 5 candidate updates.

	\begin{figure}[ht]
		\centering
		\includegraphics[width=\linewidth]{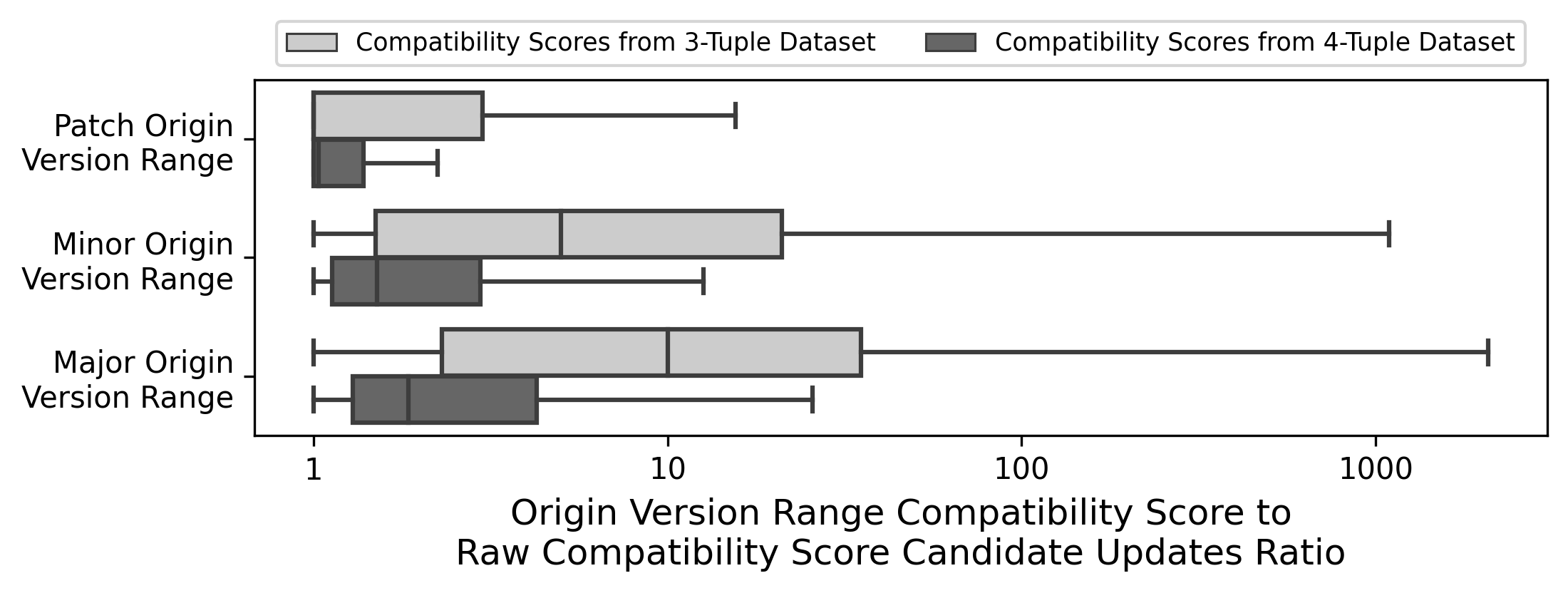}
		\caption{The distribution of ratios of candidate updates for each origin version range compatibility score to the associated raw compatibility score from the 3-tuple and 4-tuple datasets.}
		\label{\lblroot:fig:range_candidate_updates_by_dependency_update_type}
	\end{figure}

	\end{observation}

	\smallskip \begin{observation}{Both the \rangecompatibilityscores and the \clienthistorymetrics dimensions have significant predictive power to assess whether a client package developer will accept or reject a dependency update.}
	The \rangecompatibilityscores model achieves a median AUC 2.4\% higher (0.64) than the base model, with the minor origin version range compatibility score having the highest permutation importance. The \clienthistorymetrics model performs even better, achieving a median AUC 21.5\% higher (0.76), with the number of \dbprs previously merged in the client package having the highest permutation importance. Figure~\ref{\lblroot:fig:model_auc_distributions} shows the distribution of AUC improvements of both of these models compared to the baseline model median AUC. Figure~\ref{\lblroot:fig:models_permutation_importances:model_rcs_varimp_distributions} and Figure~\ref{\lblroot:fig:models_permutation_importances:model_chm_varimp_distributions} show the distribution of permutation importance's of each metric for the \rangecompatibilityscores model and the \clienthistorymetrics model, respectively.
	\end{observation}

	\newcommand{\subfigurewidth}{0.26\textwidth}

	\begin{figure*}[ht]
	\begin{subfigure}[b]{\subfigurewidth}
		\includegraphics[
			width=\linewidth,
			keepaspectratio
		]{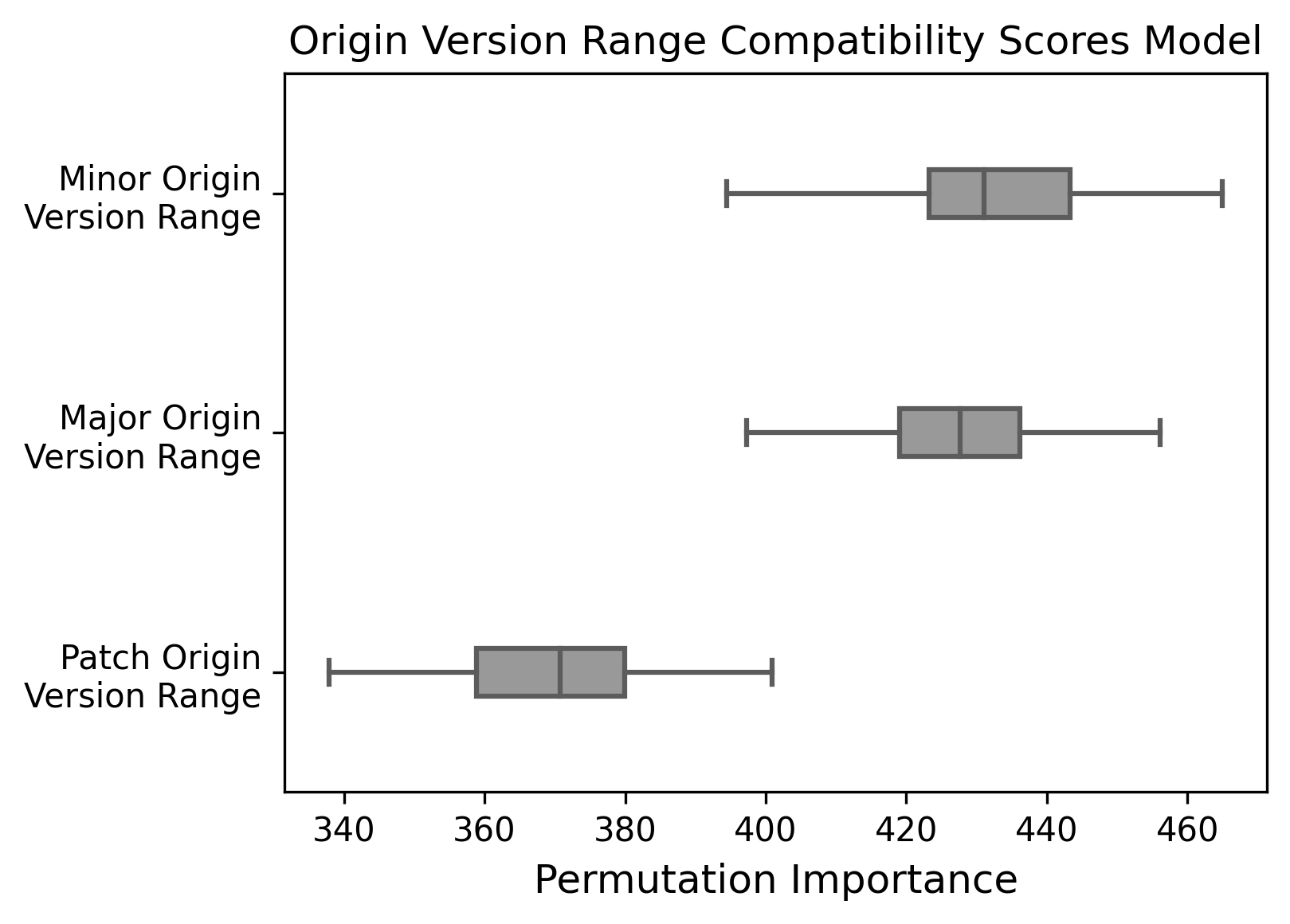}
		\caption{Distribution of permutation importance's of each metric from the \rangecompatibilityscores model.}
		\label{\lblroot:fig:models_permutation_importances:model_rcs_varimp_distributions}
	\end{subfigure}%
	\hspace*{\fill}   
	\begin{subfigure}[b]{\subfigurewidth}
		\includegraphics[
			width=\linewidth, 
			keepaspectratio
		]{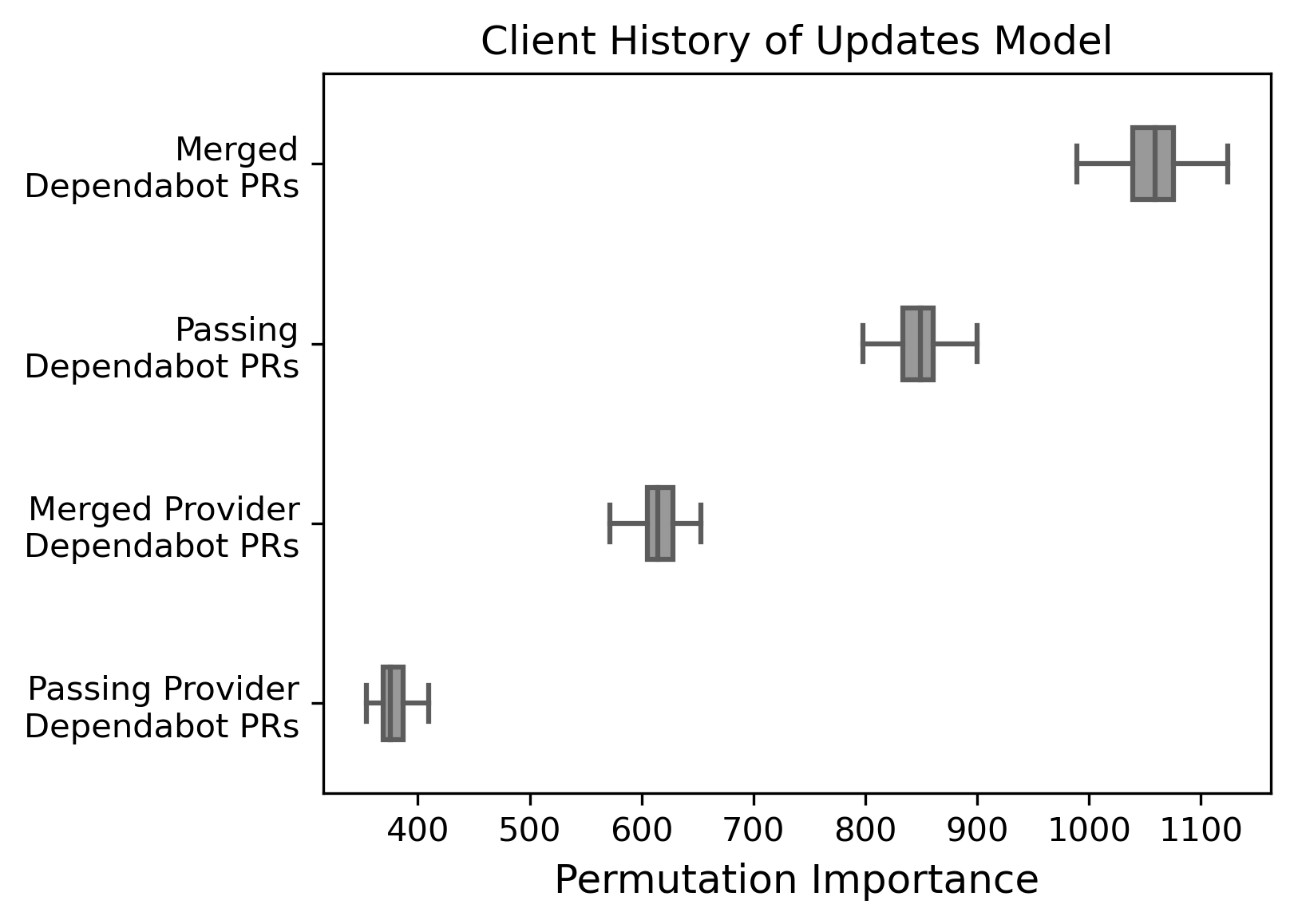}
		\caption{Distribution of permutation importance's of each metric from the \clienthistorymetrics model.}
		\label{\lblroot:fig:models_permutation_importances:model_chm_varimp_distributions}
	\end{subfigure}
	\hspace*{\fill}   
	\begin{subfigure}[b]{\subfigurewidth}
		\includegraphics[
			width=\linewidth, 
			keepaspectratio
		]{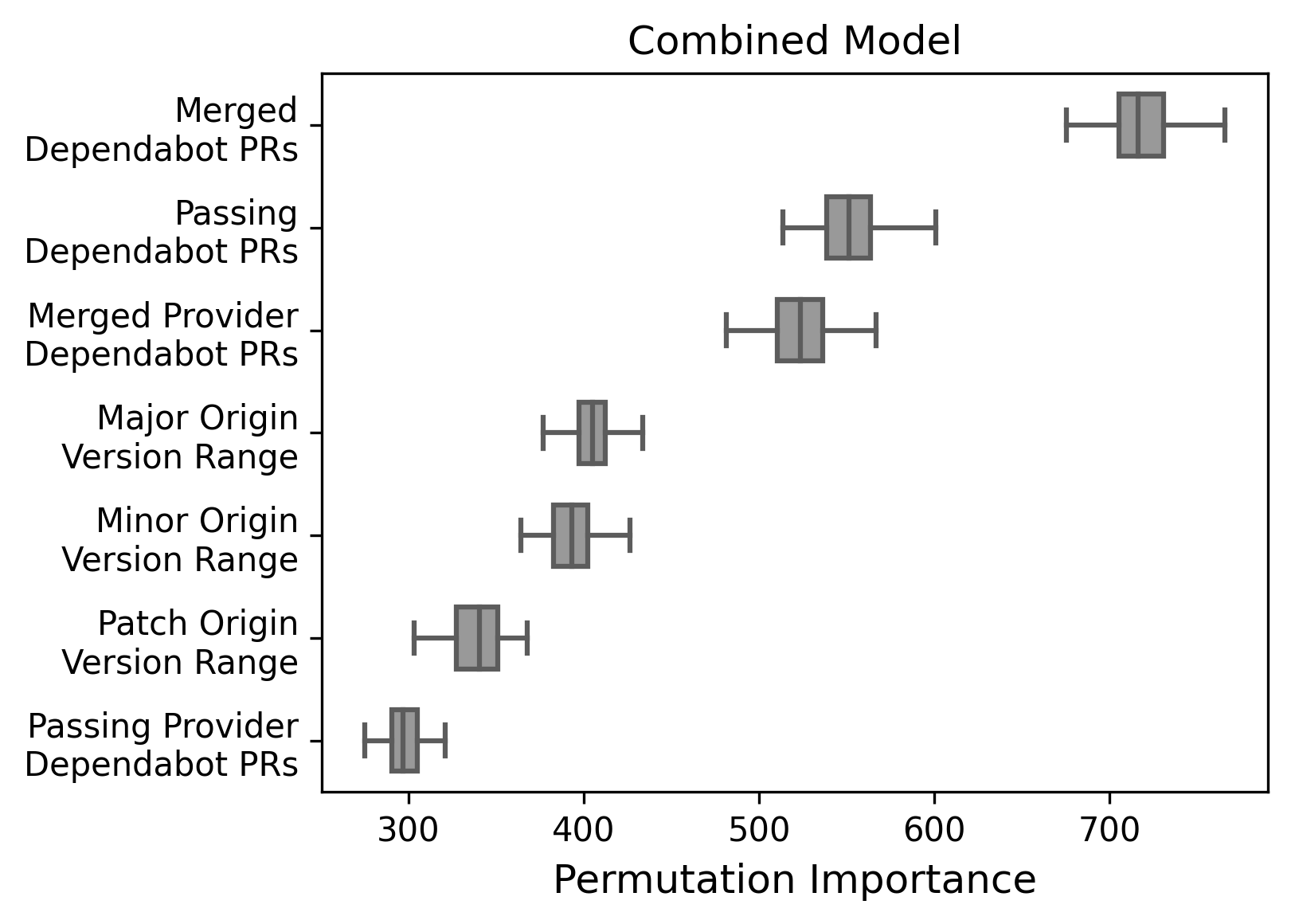}
		\caption{Distribution of permutation importance's of each metric from the model with combined metrics from both dimensions.}
		\label{\lblroot:fig:models_permutation_importances:model_both_varimp_distributions}
	\end{subfigure}
	\caption{The distribution of permutation importance's of each metric for each model.}
	\label{\lblroot:fig:models_permutation_importances}
	\end{figure*}

	\smallskip \begin{observation}{Combining metrics from both dimensions into a single model results in a larger predictive power than each of the studied dimensions individually.}
	Figure~\ref{\lblroot:fig:model_auc_distributions} shows the distribution of AUC improvements of the model that combines the metrics from both the \rangecompatibilityscores dimension and the \clienthistorymetrics dimension compared to the median AUC of the baseline model. It can be seen that the combined model has a median AUC 27.4\% higher (0.80) than the base model, which is 0.18 points higher than the base model, 0.16 points higher than the \rangecompatibilityscores model, and 0.02 points higher than the \clienthistorymetrics model. Figure~\ref{\lblroot:fig:models_permutation_importances:model_both_varimp_distributions} shows the distribution of permutation importances of each metric for the model with combined metrics from both dimensions.
	\end{observation}

	\medskip

	\begin{mybox}{\parbox{8cm}{RQ2: \rqtwo}}
		\bigskip
		\begin{itemize}[topsep = 6pt, itemsep = 3pt, label=\textbullet, wide = 0pt]
			\item Considering a range of origin versions to a specific target version helps to increase the number of candidate updates, effectively amplifying the knowledge from the crowd.
			\item Metrics from the \rangecompatibilityscores and \clienthistorymetrics dimensions can help improve the prediction of whether a dependency update will be merged by a client package developer, with the model combining all metrics having the highest performance.
			\item Historical upgrade metrics from a client package tend to have the highest predictive power when considering whether said client package will accept or reject a dependency update.
		\end{itemize}
	\end{mybox}

	\subsection{RQ3: \rqthree}
	\label{\lblroot:findings:confidence_interval}
	\noindent \textbf{Motivation.}
	As previously explained, the compatibility score for a 3-tuple dependency update is calculated as the ratio of successful updates to the total number of candidate updates. The problem here is that a compatibility score with 5 successful candidate updates (i.e., 100\%) will result in the associated dependency update appearing as more compatible than a dependency update with a compatibility score consisting of 99 successful updates and 1 failed update (i.e., 99\%). But clearly, the latter dependency update is more likely to result in further successful updates. We use the phrase \quotes{more likely} since it is possible that the dependency update with the former compatibility score consisting of 5 successful updates is in fact more compatible in other client packages than the latter with the compatibility score consisting of 99 successful updates. The hesitation to agree with this hypothesis is because we have not seen the other 95 potential candidate updates that would contribute to the compatibility score of the former dependency update. Perhaps it will achieve an additional 95 successful updates and 0 failed updates and be considered better than the latter, though not likely.

	Not only does the quantity of candidate updates affect how much confidence client packages should place in the compatibility score, but also the quality of these candidate updates. For example, if a \dbpr is contributing as a candidate update to a compatibility score, but the client's CI pipeline only consists of a linter check, it will bear the same weight as a candidate update with which the associated client's CI pipeline consisting of a build check, unit \& integration test checks, and a deployment check. Evidently, client packages would be more inclined to place a higher level of confidence (i.e., trust) in a compatibility score that was calculated using candidate updates similar to the latter example rather than the former. Therefore, in this RQ, we investigate 1) how trustworthy are the compatibility scores based on the quantity of candidate updates? and 2) how trustworthy are the compatibility scores based on the quality of candidate updates?

	\smallskip \noindent \textbf{Approach.}
	Our goal is to explore how much confidence (i.e., trust) client packages should place in the compatibility scores. Our approach takes into account the quantity of candidate updates to calculate a metric that helps calibrate the trust of client packages in compatibility scores. Our metric is a confidence interval that is based on the total number of candidate and successful dependency updates used to calculate the score, which is the ratio of successful updates to candidate updates for a dependency update. The higher the number of candidate updates that are used to calculate the compatibility score, the more confident we are in this compatibility score. We calculate a 90\% confidence interval for each compatibility score based on the approach described by \citet{davidson-pilon_bayesian_2015} (further details are given in Appendix~\ref{\lblroot:appendix:ci_calculation}). We choose a 90\% confidence level because it is the least strict of the three most commonly used confidence levels (i.e., 90\%, 95\%, and 99\%), as we aim to help clients estimate the compatibility of a dependency update, which does not require exact measurements as is the case in other, more critical situations (e.g., dealing with human life)~\citep{hazra_using_2017}.

	To explore the trustworthiness of the compatibility scores from the standpoint of quantity of candidate updates, we examine the distribution of confidence intervals across both the 3-tuple and 4-tuple datasets, as well as the distance from the compatibility score to the furthest confidence interval bound (i.e., the confidence interval precision). To explore the trustworthiness of the compatibility scores from the standpoint of quality of candidate updates, we examine the number and types of checks that ran against \dbprs in our 4-tuple dataset, which were classified in Section~\ref{\lblroot:data:dependabot_pull_requests}. We exclusively examine \dbprs that had a previously passing test suite (i.e., the CI pipeline has a successful conclusion on the main branch the \dbpr is based off), which is a requirement for \db to consider the \pr as a candidate update.

	\smallskip \noindent \textbf{Findings.}
	\begin{observation}{The confidence Dependabot has in compatibility scores can vary wildly, even though they are always presented to client packages as a similar badge.} 
	We find that half of compatibility scores with at least 5 candidate updates have a 90\% confidence interval precision (i.e., the distance from the compatibility score to the furthest CI bound, see Appendix~\ref{\lblroot:appendix:ci_calculation} - Equation~\ref{\lblroot:eq:score_ci_precision}) greater than 15\%. The precision improves when examining compatibility scores from the 4-tuple dataset, with the median confidence interval precision dropping to 3.5\%. This improvement is expected, as compatibility scores from the 4-tuple dataset tend to have more candidate updates than those in the 3-tuple dataset. This shows that, while \db will present every compatibility score as a similar badge on \prs, it is very common for the confidence \db has in these scores to vary wildly. This can mislead client packages into thinking the dependency update is in fact more stable than it actually is. The distributions of the 90\% confidence interval precision for both the 3-tuple and 4-tuple datasets are shown in Figure~\ref{\lblroot:fig:ci_precision_distribution_at_least_5_split}.

	\begin{figure}[ht]
		\centering
		\includegraphics[width=\linewidth]{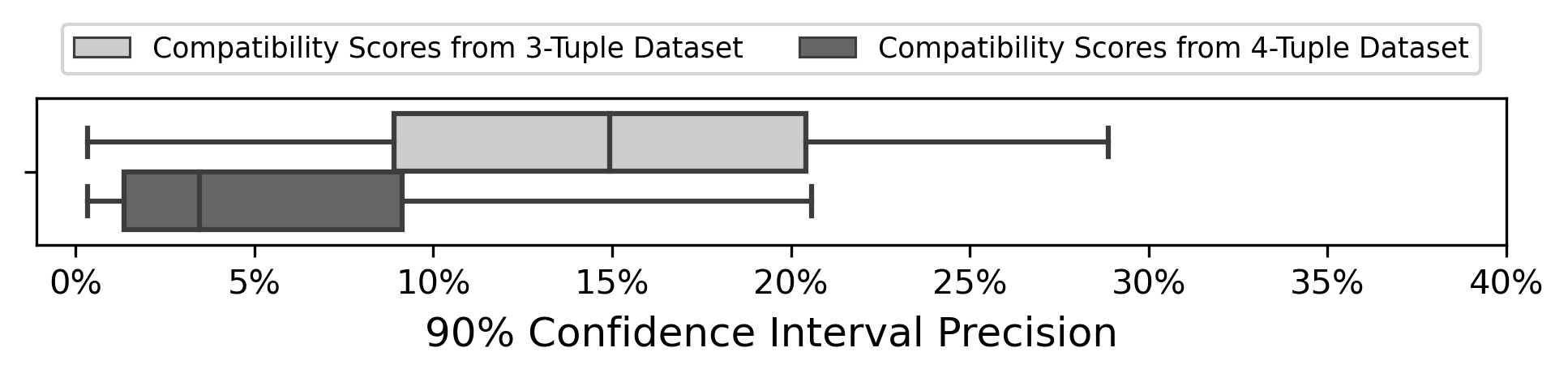}
		\caption{The distributions of the 90\% confidence interval precision for both the 3-tuple and 4-tuple datasets.}
		\label{\lblroot:fig:ci_precision_distribution_at_least_5_split}
	\end{figure}

	\end{observation}

	\begin{observation}{CI pipelines for candidate update Dependabot PRs often contain a mixture of check types that are not always helpful for testing the compatibility of the dependency update.}
	We find that candidate update \dbprs have a median of 3 checks that make up the client's CI pipeline to test the dependency update. The vast majority (94\%) have at least a build or test check that is part of the CI pipeline. However, while these checks may seem promising, it is worth noting that it is common for client's tests to not thoroughly exercise the package's dependencies~\citep{hejderup_can_2021}. Additionally, recall that just over 1 in 10 (11\%) check runs we collected were considered to be useless from the standpoint of contributing to the compatibility score. A quarter (26\%) of candidate update \dbprs have a CI pipeline that contains at least one of these useless check, while 1\% candidate update \dbprs have only useless checks that ran against the dependency update. Of the candidate update \dbprs with only useless checks, 94\% of them had a successful CI conclusion, compared with 88\% of \dbprs with at least one build check.

	\end{observation}

	\begin{mybox}{\parbox{8cm}{RQ3: \rqthree}}
		\medskip
		\begin{itemize}[topsep = 6pt, itemsep = 3pt, label=\textbullet, wide = 0pt]
			\item{Client packages should be hesitant to place total confidence in the accuracy of compatibility scores, as more than half of the scores with at least 5 candidate updates have a 90\% confidence interval precision greater than 15\%.}
			\item{Candidate updates that contribute to compatibility scores may not always truly test the associated dependency update.}
		\end{itemize}
	\end{mybox}

	\section{Discussion}
	\label{\lblroot:discussion}
	In this section, we discuss the findings observed in Section~\ref{\lblroot:findings}. We present a set of practical implications for designers of dependency management bots with the aim of using the crowd to help client packages assess the risk of accepting a dependency update.

	\smallskip \implication{When the crowd does not provide enough support to calculate a compatibility score for a specific dependency update, dependency management bots should consider candidate updates from different origin version ranges to amplify input from the crowd.}
	We found in RQ1 that fewer than 1 in 5 of dependency updates have at least 5 candidate updates, which is the threshold required for the compatibility score to be shown on \dbprs. We hypothesize that such a low proportion of compatibility scores with less than 5 candidate updates can be partially explained by two reasons. First, while \db may be opening a high number of \prs in client packages for dependency updates, only a small portion of these client packages actually meet the requirements set by \db for these \prs to be considered as candidate updates that contribute to the associated compatibility score (i.e., the client has a CI pipeline configured and a previously passing test suite on the main branch). Second, while many client packages may use the same provider package as a dependency, \db may not be creating \prs to update that provider from the same origin and target versions. For example, for the provider package $P$ and the 3 versions $V_1$, $V_2$, and $V_3$ of $P$, the 3-tuple updates ($P$, $V_1$, $V_3$) and ($P$, $V_2$, $V_3$) will have two different compatibility scores with separate candidate updates. So, while there may be a potentially high number of candidate updates for the provider package overall, all of these candidate updates end up being spread across a wide range of potential origin version and target version combinations, resulting in a low number of candidate updates for each specific origin version and target version combination.

	While exploring the idea of \db considering candidate updates from a range of origin versions in RQ2, we found that this ratio increases to 2 in 5 when considering all candidate updates from dependency releases with only a different patch origin version number, and to over two-thirds when considering all candidate updates from dependency releases with only a different minor or patch origin version number. Moreover, while considering the candidate updates from the patch origin version range did not result in a significant increase for the majority of compatibility scores, we found that considering a minor origin version range resulted in 5x the number of candidate updates as the raw compatibility score, while considering a major origin version range resulted in 10x the number of candidate updates as the raw compatibility score. This is reflected in our results from RQ2, where we found that the minor and major origin version range compatibility scores have the highest permutation importance in the origin version range compatibility scores model.

	These are significant improvements which can lead to a more general form of the compatibility score being available and useful to a higher number of client packages while attempting to minimize the accuracy lost due to the range of origin versions being considered for the score. The \semver policy specifies that important backward compatible changes require an update of the minor version component, and backward compatible bug fixes require an update of the patch version component~\citep{decan_what_2020}. Assuming this policy is followed by provider package maintainers, which \citet{decan_what_2020} found is becoming more common as software ecosystems mature, considering patch and minor origin version ranges will still be able to provide a reasonably accurate compatibility score, as major origin version range compatibility scores may be biased by breaking updates purposely introduced by package maintainers, since the \semver scheme specifies that backward incompatible changes require an update of the major version component. However, designers of dependency management bots should make it clear to client packages that the origin version range compatibility scores might not be representative of the exact dependency update the client package is considering.

	\smallskip \implication{When there are simply not enough candidate updates from the crowd, dependency management bots should consider historical update metrics from the client package.}
	We found in RQ1 that 83\% of compatibility scores do not have enough candidate updates to be shown on \dbprs. We explored potential solutions to this issue in RQ2, one of which was to consider the candidate updates from a range of origin versions, which we discussed in the previous implication. However, there are still cases where there are simply not enough candidate updates from the crowd to calculate a trustworthy compatibility score, even when considering the origin version range scores. Specifically, more than half (61\%) of patch origin version range compatibility scores still have fewer than 5 candidate updates.

	In these situations, dependency management bots should turn to historical client upgrade metrics to help clients assess whether they should accept or reject a dependency update. Not only did this dimension result in the model with the highest median AUC (improvement of 21.5\% over the baseline), but we also found that the number of \dbprs previously merged by a client package developer and the number of \dbprs previously passing the client's CI pipeline were the most important features in our combined model. This suggests that taking historical upgrade metrics from the client package into account when considering a dependency update could be an effective way to provide a personalized compatibility score for each client.

	In fact, it can be beneficial for dependency management bots to take these additional metrics into account not only when input from the crowd is low, but also when input from the crowd is high. We tested our models on \dbprs with at least 5 candidate updates, and found similar results as in RQ2, with the combined model achieving an AUC of 0.78, 0.16 points higher than the baseline model.

	\smallskip \implication{Regardless of the level of input from the crowd, dependency management bots should provide supporting metrics alongside compatibility scores to signal the level of trust client packages should place in the compatibility score.}
	We found in RQ1 that it is common for compatibility scores to have a low number of candidate updates, and what is lacking from the compatibility score is supporting information that tells client packages how much they can trust the recommendation from \db. When people interact with any complex system (e.g., software bots), they create a mental model, which facilitates their use of the system~\citep{norman_design_2002, kulesza_tell_2012}. In automation-supported software engineering (e.g., deciding whether to update a dependency), valid mental models of the reliability of the output (e.g., the compatibility score of the dependency update) help the user (e.g., a client package) to know when to trust the recommendation. In fact, \citet{zhang_effect_2020} found that confidence intervals can help calibrate people's trust in automation-supported decision making. Similarly, confidence intervals could help \db's compatibility score by providing client packages with an estimate of its trustworthiness, so that clients are able to distinguish between a 100\% compatibility score with only 5 candidate updates and a 99\% compatibility score with 99 candidate updates.

	We explore this idea of using a confidence interval to help calibrate the level of trust client packages should place in the compatibility score in RQ3. We find that if a 90\% confidence interval based on the number of candidate updates used to calculate a compatibility score was included on \dbprs, half would show that the confidence interval precision was greater than 15\%. These findings suggest that the level of trust client packages should place in compatibility scores can vary wildly, even though the compatibility score is always presented as the same badge style on \dbprs. Therefore, dependency management bots should include additional metrics, like the confidence interval we calculated in RQ3, that can help to calibrate the level of trust client packages should place in the compatibility score. Presenting this information could be especially useful to client packages that do not have a CI pipeline configured, as they stand to gain the most benefit out of leveraging the crowd to assess the risk of a dependency update, and therefore should be aware of the level of confidence with which the associated compatibility score has been calculated.

	\smallskip \implication{Regardless of the level of input from the crowd, dependency management bots should place higher weights on packages with high quality CI pipelines that thoroughly test the dependency being updated.}
	We found in RQ3 that candidate updates that contribute to compatibility scores can have CI pipelines that contain a variety of different types of checks, and may not always test the associated dependency update. In the extreme scenario, we found that of the 1\% of \dbprs that only had useless checks run against the dependency update, 94\% of them had a successful CI conclusion. So, while these \dbprs may be contributing as successful updates towards the compatibility score, they have not tested the dependency at all.

	To address this issue, dependency management bots should take into account additional metrics other than the simple pass/fail result of a client's CI pipeline, such as the types of tests and the level of test coverage in the client package, to ensure that only high-quality input from the crowd is being considered. We saw in RQ1 that compatibility scores are already heavily skewed toward the higher range, which may be influenced by the fact that too many of the candidate updates contributing to the scores are from low-quality pipelines in client packages that do not truly test the dependency update. Therefore, dependency management bots should attempt to quantify the level of quality of clients from the crowd, and then either only consider clients that truly test the dependency update (i.e., have a high-quality CI pipeline), or perhaps provide a weight to each client based on the level of quality of their CI pipeline. This idea is similar to that of \quotes{Security Scorecards}\fnu{https://github.com/ossf/scorecard}, in which a number of heuristics associated with software security are tested against a package's dependencies and assigned a score of 0-10. Dependency management bots could apply the same principle against a package's CI pipeline, evaluating heuristics related to software build and test quality in order to assess whether they should consider the CI pipeline when evaluating the compatibility of a new provider package release. However, designers of dependency management bots should be mindful that, while this may lead to a higher quality compatibility score, the trade-off is that fewer candidate updates may be available to calculate the compatibility score.
		
	\section{Related Work}
	\label{\lblroot:related_work}
	In this section, we discuss related work to automated dependency management (Section~\ref{\lblroot:related_work:automated_dependency_management}) and crowd-sourced software engineering (Section~\ref{\lblroot:related_work:crowdsourcing}).

	\subsection{Automated Dependency Management}
	\label{\lblroot:related_work:automated_dependency_management}
	Multiple studies examine the growing trend of using third-party provider packages to build new software~\citep{wittern_look_2016, fard_javascript_2017}, sometimes even for trivial tasks~\citep{abdalkareem_why_2017, abdalkareem_impact_2020, chowdhury_untriviality_2021}. However, multiple studies~\citep{decan_evolution_2018, gonzalez-barahona_technical_2017, zerouali_empirical_2018} find that clients are reluctant to update their dependencies and that the difference between the client's currently used version of a provider and the provider's latest release (i.e., \textit{technical lag}) is increasing over time. Existing research finds that this can lead to security issues~\citep{decan_impact_2018, cox_measuring_2015}, depending on deprecated releases~\citep{cogo_deprecation_2021}, and dependency conflicts~\citep{decan_empirical_2017}.

	\citet{bogart_when_2015} find that the risk of breaking changes is one of the primary concerns that developers experience when it comes to dependency management, and multiple works study how to detect breaking changes in dependency updates~\citep{brito_apidiff_2018, moller_model-based_2019}. Most relevant to our paper, \citet{hejderup_can_2021} find that client's tests can only detect on average 47\% of direct and 35\% of indirect artificial faults and note that a combination of static and dynamic analysis should be used to effectively detect breaking updates. 

	Automated dependency management bots have become a popular area of research~\citep{erlenhov_current_2019, lebeuf_defining_2019, wessel_enhancing_2020, wessel_inconvenient_2020}. Specifically, ~\citet{mirhosseini_can_2017} find that clients who use dependency bots tend to upgrade their dependencies 1.6 times as often clients that do not. In the same vein, \citet{alfadel_use_2021} find that 65\% of \db security \prs are merged and integrated in the packages, often within a day, and that the large majority of unmerged \db security \prs (94\%) were closed by \db itself because the \pr had become outdated.

	\subsection{Crowd-sourced Software Engineering}
	\label{\lblroot:related_work:crowdsourcing}
	Several researchers have studied how to use the \quotes{wisdom of the crowds} to help in the software engineering domain. \textsf{Stack Overflow}\fnu{https://stackoverflow.com/}, a popular Question and Answer (Q\&A) site, has been the topic of many studies~\citep{latoza_crowdsourcing_2016, treude_how_2011, rosen_what_2015, barua_what_2014, vasilescu_stackoverflow_2013} that explore how the job of answering questions related to software engineering is outsourced to the crowd. Specifically, \citet{abdalkareem_what_2017} analyze 1,414 \textsf{Stack Overflow} related commits and observe that developers use this crowd based knowledge mostly for technical comprehension, collecting users’ feedback and code reuse.

	The idea of using the crowd to help with dependency management has also been studied. \citet{mileva_mining_2009} proposed an approach and associated tool to help client packages decide when to use which version of a provider package. They reason that if a provider package version is used by more client packages, it should be more likely to be recommended. To use the crowd to detect breakage changes in new provider releases,~\citet{mujahid_using_2020} propose a technique that leverages the automated test suites of other client packages that make use of the same dependency to test newly released versions, describing essentially an academic version of the \db compatibility score. They find that this crowd-based approach can detect six of ten breakage-inducing versions they studied, and that their findings can help clients to make more informed decisions when they update their dependencies, which is also the goal of the \db compatibility score. Similarly,~\citet{mezzetti_type_2018} describe an approach called \textit{type regression testing} to automatically detect type-related breaking changes, This approach leverages the tests of clients of a provider package construct models of new releases of provider package APIs, and then compare these models to detect potential \textit{type regressions}, demonstrating that this approach can detect type-related breaking changes with high accuracy. They argue that using the clients' test suite, rather than the test suite of the provider package itself, are more likely to provide representative executions and only use the public parts of the provider package.

	\section{Threats to Validity}
	\label{\lblroot:threats}
	In this section, we discuss the threats to the validity of our study.

	\subsection{Internal Validity}
	\label{\lblroot:threats:internal}
	Threats to internal validity concerns factors that could have influenced our analysis and findings. The \db API simply returns every compatibility scores \db has a record of at the time of the request. This means that we only retrieve a single snapshot of the compatibility scores for the dependencies being updated, and that the scores we collected for specific updates might not be the actual scores that a practitioner saw when \db opened a \pr on their package. To investigate how this might have affected our analysis, we built a pipeline that runs three times per day (i.e., every 8 hours), and retrieves all the \dbprs for the list of packages described in Section~\ref{\lblroot:data:project_selection} that have been created since the previous pipeline run. We then extract the 3-tuple data for the provider package being updated, and retrieve the compatibility score for that update every time the pipeline runs for the next 14 days. If a different client has a \dbpr for that same 3-tuple update, the 14 day threshold will reset. We let the pipeline run from November 25, 2020 until April 1, 2021. This gives us a time series dataset of compatibility scores that allows us to explore how long it might take or how many candidate updates are required for a compatibility score to become stable. We consider a score to have become stable at the latest point in time where the score is within 5\% of the final score record. We find that 85.6\% of compatibility scores we collect have the first score and all subsequent scores varying within 5\% of the final compatibility score. In other words, only 14.4\% of the compatibility scores were not instantly stable, which suggests that the analysis we conducted in our study would not have been significantly impacted by the fact that we only collect a single snapshot of the compatibility scores.

	Additionally, the \db API will not return a record for a specific 3-tuple dependency update if \db has not recorded any candidate updates for said 3-tuple. Consequently, all compatibility scores we analyze have at least 1 candidate update. This means that the proportion of dependency updates that have compatibility scores with at least 5 candidate updates is likely lower than what we observe, a point which we mention in RQ1, as we disregard any dependency updates that have 0 candidate updates.

	Another concern is related to the conclusions drawn when we consider whether or not a dependency update on a \dbpr caused the client's CI pipeline to fail. We use the checks associated with each \pr to determine whether the update failed the CI pipeline, but checks can fail for reasons unrelated to the dependency update (e.g., flaky tests, license issues, etc.). However, this would have minimal effect in the context of studying the compatibility score, as \db does not take the type of check that failed on a \dbpr into account when considering the \pr as a candidate update that contributes to the compatibility score.

	Finally, it is important to note that the conclusions drawn when considering whether or not a client package merged a \dbpr may have been affected by the fact that factors other than the compatibility score can contribute to the client package's decision of whether to merge the \dbpr or not. For example, a client package may be very risk averse, resulting in very few merged \dbprs, regardless of the compatibility signals provided by \db. Still, we attempt to account for this threat by taking into account the historical upgrade metrics for each client package when performing this analysis. 

	\subsection{External Validity}
	\label{\lblroot:threats:external}
	Threats to external validity concern the generalization of our technique and findings. Our study only focuses on the compatibility score implementation of \db. Therefore, our results cannot be generalized to different implementation of leveraging knowledge from the crowd to provide insights about the risk of a newly released version of a provider package. For example, the \textsf{Renovate} bot has a feature similar to \db's compatibility score called \textit{Merge Confidence}\fnu{https://docs.renovatebot.com/merge-confidence/} that identifies and flags undeclared breaking releases based on analysis of test and release adoption data. \textsf{Renovate}'s \textit{Merge Confidence} has unique features that might differ from our results with \db's compatibility score. Still, to the best of our knowledge, \db was the first to leverage the crowd for dependency management by providing a compatibility score for each dependency update. So, while our results cannot be generalized, our discussion provides implications that can still apply to other bots that leverage the crowd to assess the risk of an update.

	\section{Conclusion}
	\label{\lblroot:conclusion}
	Today's software systems are rarely built from scratch, with client packages often making use of specific versions of provider packages in the form of dependency relationships. These dependency relationships come with the essential and risky task of keeping the client package's dependencies up-to-date. \db, an automated dependency management bot, helps facilitate this process by automatically opening \prs in client packages to update a dependency when a new version of the provider is released, as well as providing a \textit{compatibility score} for each dependency update. This compatibility score is shown as a badge on \prs opened by \db, and is meant to give clients a sense of the risk involved when updating a dependency by leveraging the knowledge of \quotes{the crowd}, so that clients can be confident a new provider version is backwards compatible and bug-free.

	In this paper, we perform an empirical study of 579,206 \dbprs, as well as 618,045 compatibility score records, that examines the viability of dependency management bots leveraging the crowd to help clients assess the risks involved with accepting a dependency update. 
	We conclude that dependency management bots should go beyond only considering the result of clients' CI pipeline running against a dependency update when using the crowd to assess the risk of said dependency update. 
	This is especially relevant since we found that the majority of compatibility scores do not have at least 5 candidate updates, which is the threshold required for the compatibility score badge to be displayed on \dbprs, and when compatibility scores do have enough candidate updates, the vast majority of the scores are above 90\%. 
	As a result of this skewness in both the number of candidate updates and the scores themselves, dependency management bots should employ further methods to help amplify input from the crowd or consider historical upgrade metrics to assess whether a client package should accept or reject a dependency update.
	Additionally, supporting metrics, such as a confidence interval, should be provided alongside the compatibility score to help calibrate the level of trust client packages should place in the score. 

    \section{Acknowledgment}
	\label{\lblroot:sec:acknowledgements}
	The findings and opinions in this paper belong solely to the authors, and are not necessarily those of Huawei. Moreover, our results do not in any way reflect the quality of Huawei software products.

    \bibliographystyle{BensIEEEtranN}
    \bibliography{dbcs_references.bib}
	\newpage


	\begin{IEEEbiography}
		[{\includegraphics[width=1in,height=1.25in,clip,keepaspectratio]{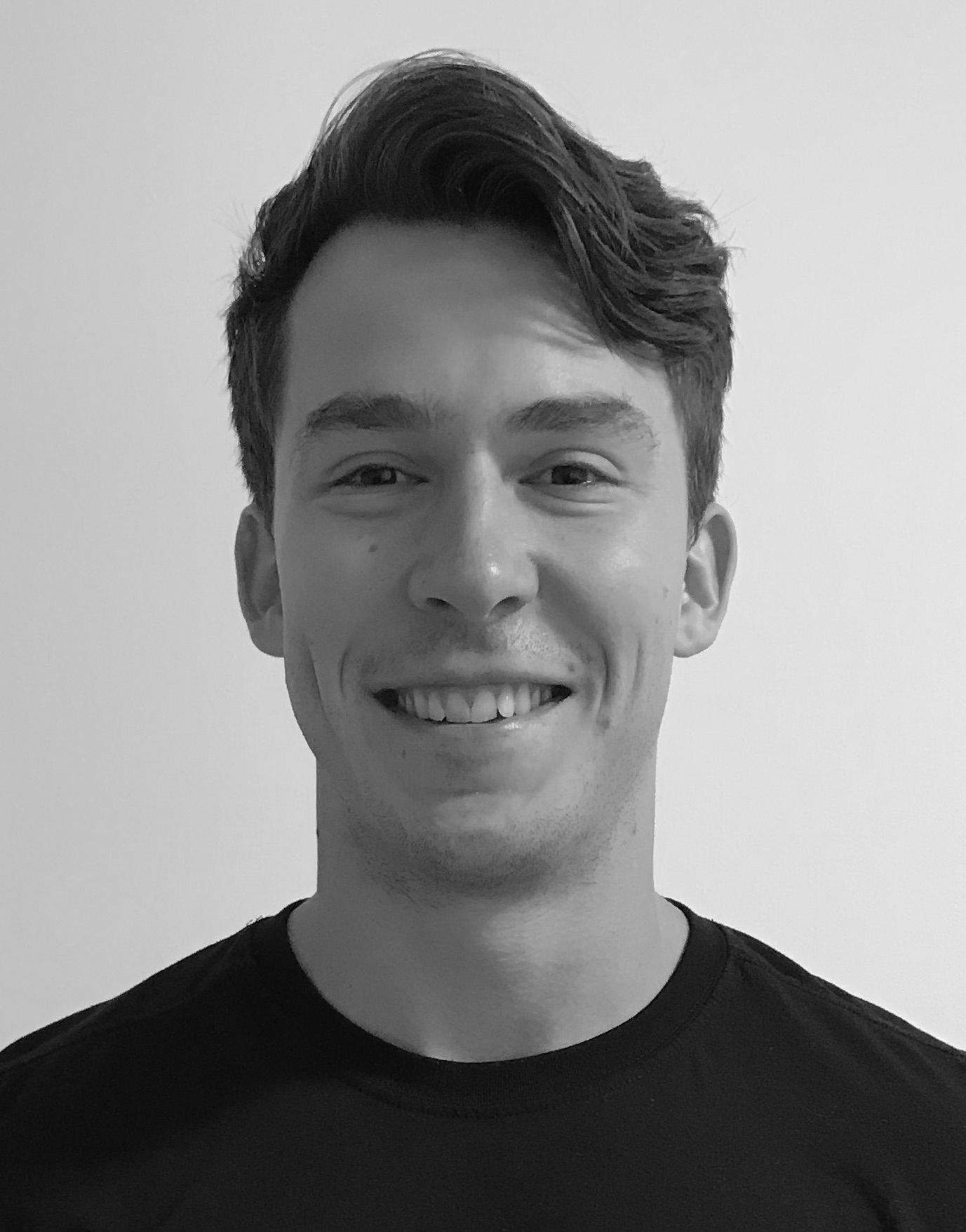}}]
		{Benjamin Rombaut} is a MSc candidate with the Software Analysis \& Intelligence Lab (SAIL) at Queen's University, Canada, as well as a Software Engineering Researcher at Huawei, Canada. He received his BSc in Software Engineering from the University of New Brunswick, Canada in 2019. His research interests include empirical software engineering, software supply chain management, and software bots. More information at \mbox{\url{https://www.benrombaut.ca}}.
	\end{IEEEbiography}

	\vskip -2\baselineskip plus -1fil

	\begin{IEEEbiography}
		[{\includegraphics[width=1in,height=1.25in,clip,keepaspectratio]{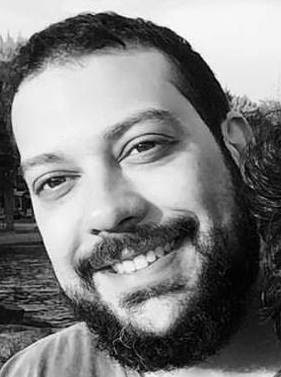}}]
		{Filipe R. Cogo} is a Software Engineering Researcher at Huawei, Canada. His research focuses on empirical software engineering and mining software repositories. He received his BSc and MSc in Computer Science from Universidade Estadual de Maringá (UEM), Brazil, and his PhD from Queen’s University, Canada.
	\end{IEEEbiography}

	\vskip -2\baselineskip plus -1fil

	\begin{IEEEbiography}
		[{\includegraphics[width=1in,height=1.25in,clip,keepaspectratio]{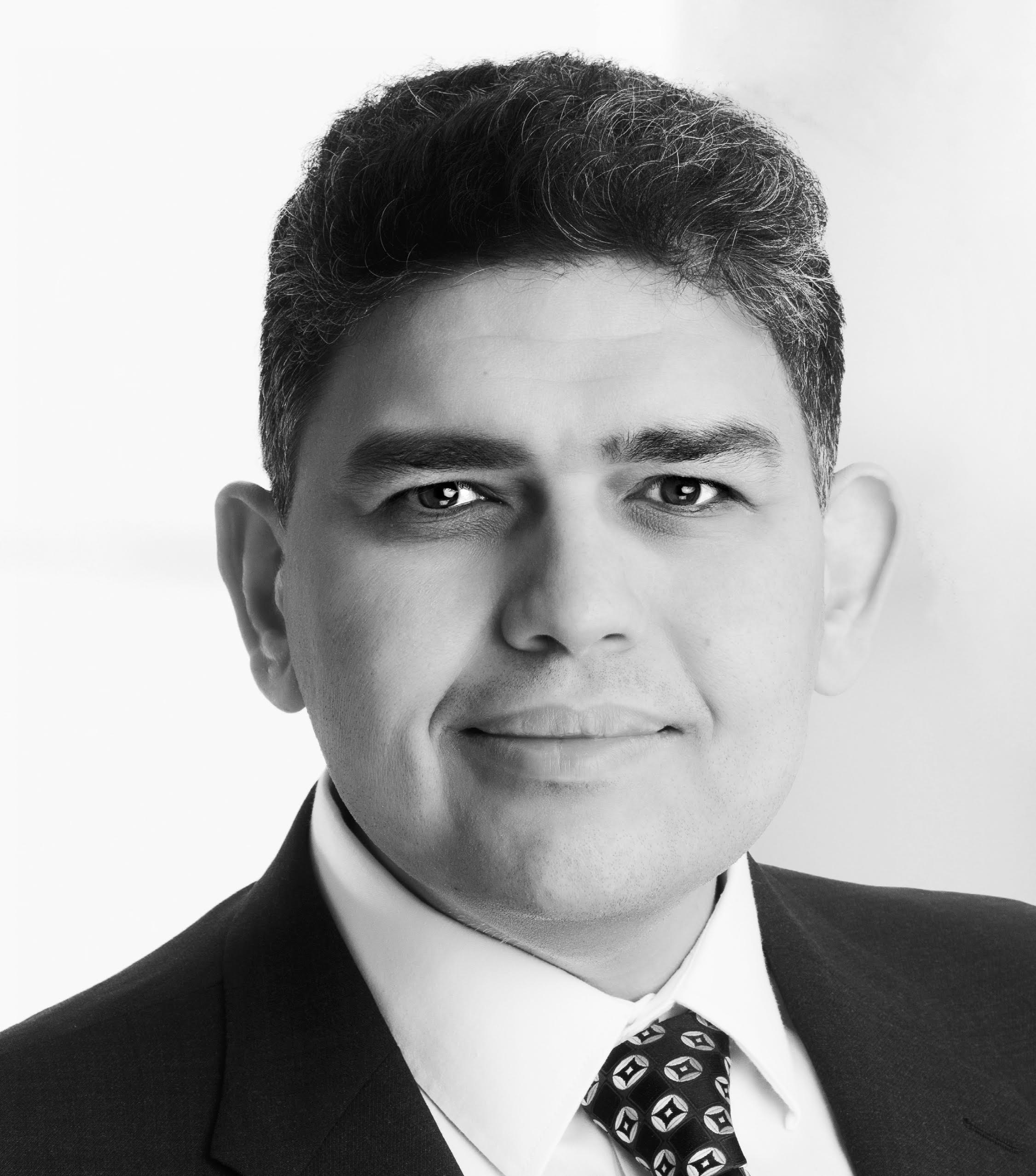}}]
		{Ahmed E. Hassan} is an IEEE Fellow, an ACM SIGSOFT Influential Educator, an NSERC Steacie Fellow, the Canada Research Chair (CRC) in Software Analytics, and the NSERC/BlackBerry Software Engineering Chair at the School of Computing at Queen’s University, Canada. His research interests include mining software repositories, empirical software engineering, load testing, and log mining. He received a PhD in Computer Science from the University of Waterloo. He spearheaded the creation of the Mining Software Repositories (MSR) conference and its research community. He also serves/d on the editorial boards of IEEE Transactions on Software Engineering, Springer Journal of Empirical Software Engineering, and PeerJ Computer Science. Contact ahmed@cs.queensu.ca. More information at: \url{http://sail.cs.queensu.ca}
	\end{IEEEbiography}

	\newpage
	\begin{appendices}
	    \section{Patterns for classifying CI build types}
		\label{\lblroot:appendix:ci_name_regex}
		The following steps are used to classify the checks that ran as part of the client's CI pipeline on \dbprs: 
		i) the first author manually examined the 20 most popular unclassified check names (a check name is used to give a high-level description of the task the check performs) to extract patterns that could be grouped into similar categories, 
		ii) these new patterns are added to a set of regular expressions that capture common check names and assign these checks to a specific overarching check type category, 
		iii) the full data set of checks are then re-classified with the updated regular expressions, 
		iv) the process is repeated using only checks that have not yet been classified until any new extracted patterns do not classify a threshold of at least 0.01\% of the unclassified checks. Once this threshold was reached, 91.8\% of the checks had been matched to one of six categories described in Table~\ref{\lblroot:table:ci_name_regex}.

		\begin{table}[ht]
			\caption{
			String patterns for classifying the types of checks that ran against \dbprs.
			}
			\begin{center}
			\begin{tabularx}{\linewidth}{>{\hsize=.17\hsize}X>{\hsize=.15\hsize}XX}
			\toprule
			\textbf{Category} & \textbf{Percent} & \textbf{Regular Expression\newline (case insensitive)} \\
			\toprule
			Build & 58.1\% & \code{(\textasciicircum | |-)(build|install)|Travis CI|developing-with-angular|\newline (main|workflow|setup)|Node(.js)? \textbackslash \textbackslash d?\textbackslash \textbackslash d?|(Continuous integration|\textasciicircum ci(\$| ))|(tsc|typescript)|monica CI|(web|webpack)|PHP|(Try: )?ember((-| )try)?|\newline (macOS|windows|ubuntu|linux)\newline (-latest)?|Python|\textasciicircum 3.\textbackslash\textbackslash d\$|\textasciicircum 2.\textbackslash\textbackslash d\$} \\
			\midrule
			Test & 17.2\% & \code{(\textasciicircum| )test|Analy(s|z)e|Analysis|\newline karma|e2e| stoplightio|check|\newline~unit-js| run|Validation|rspec} \\
			\midrule
			Useless & 11.2\% & \code{WIP|\textasciicircum Rule: automatic merge for Dependabot pull requests \textbackslash(merge\textbackslash)\$|(Auto ?)?merge|\textasciicircum stale\$|\textasciicircum Update \textbackslash.NET SDK\$|\textasciicircum Summary\$|fixupbot|Mixed content|Rebase|Autosquash|Backport|\newline docs|hyperjump|kodiakhq: status|DCO|lock|Discord Listener|Label|css|Clean GitHub pages|pre-commit|remove-pr|\newline markdown-link-check|Run CircleCI artifacts redirector|pedrolamas.com|Auto Approve a PR by dependabot|dependabolt|\newline github/dependabot.yml|greeting|\newline chrome|firefox|finish|\newline mui-org.material-ui| jbhannah.net|Always run job|jhipster.generator-jhipster|\newline dispatch|Timeline protection|Inclusive Language|mark-duplicate|migration|\newline Generate HTML log|feature flags} \\
			\midrule
			Lint & 7.0\% & \code{(es)?lint|ESLint Report Analysis|codecov|Floating Dependencies|prettier| Coverage|Standard| bundle-size|pronto|flake8| mypy|CodeFactor|Code style} \\
			\midrule
			Deploy & 4.9\% & \code{Redirect rules|Header rules|deploy|release|Pages changed|publish|artifact} \\
			\midrule
			Security Analysis & 1.6\% & \code{code(| |-)ql|GitGuardian Security Checks|SonarCloud Code Analysis|LGTM analysis|depcheck|audit| rubocop} \\
			\bottomrule
			\end{tabularx}
			\end{center}
			\label{\lblroot:table:ci_name_regex} 
		\end{table}

		\section{Distinguishing between the 3-tuple and 4-tuple datasets}
		\label{\lblroot:appendix:tuple_datasets}

		The list of client packages (Section~\ref{\lblroot:data:project_selection}) that we collected \dbprs for (Section~\ref{\lblroot:data:dependabot_pull_requests}) could have resulted in having multiple \dbprs for the same 3-tuple update. For example, for the 3-tuple update ($P$, $V_1$, $V_2$) for updating provider package $P$ from version $V_1$ to $V_2$, we might have ($C_1$, $P$, $V_1$, $V_2$) and ($C_2$, $P$, $V_1$, $V_2$), where $C_1$ and $C_2$ are different client packages that make use of $P$ as a dependency. However, we might not necessarily have at least one matching 4-tuple update for every 3-tuple update. In other words, our list of client packages built in Section~\ref{\lblroot:data:project_selection} is by no means an exhaustive list of all packages that use \db. So, if \db opens a \pr in client $C$ for a 3-tuple update that is contributing as a candidate update, we would only potentially find a matching 4-tuple update for the associated 3-tuple update if $C$ is in our list of client packages.

		If client $C$ is in our list of packages, and they use provider package $P$ as a dependency, there is no guarantee that \db has opened a \pr for every 4-tuple combination of ($C$, $P$, $V_X$, $V_Y$), where $V_X$ and $V_Y$ are two versions of $P$, with $V_X$ being released prior to $V_Y$. For example, if $C$ only adopts $P$ as a dependency beginning at version $V_4$, then \db would not have opened any \prs for the 4-tuple update ($C$, $P$, $V_{X<4}$, $V_{Y<4}$) where $V_{X<4}$ and $V_{Y<4}$ are versions of $P$ that were released prior to $V_4$.

		Additionally, \db is constrained to opening \prs for dependency updates that fall within the client's dependency version specifications. For example, if $C$ has version pinned $P$ to version $V_1$, \db would only open \prs that follow the 4-tuple ($C$, $P$, $V_1$, $V_T$), where $V_T$ is the latest target release of $P$ (assuming that $C$ does not change their version specifications for $P$). 

		A final reason that might explain why \db might not open a \pr for specific 4-tuple updates is that clients are able to configure a limit for the number of \dbprs open in their package at any one time\fnu{https://docs.github.com/en/code-security/supply-chain-security/keeping-your-dependencies-updated-automatically/configuration-options-for-dependency-updates\#open-pull-requests-limit}. So if the limit of \prs allowed to be opened by \db at one time in client $C$ has already been reached (e.g., updates for other dependencies), \db will not create any new \prs to update $P$, even as $P$ releases new versions, until the number of open \dbprs in $C$ has dropped below the limit.

		As a concrete example to illustrate the distinction between the 3-tuple update and 4-tuple update datasets, we again use the scenario from Figure~\ref{\lblroot:fig:dbpr_example}, which shows a \dbpr for the 4-tuple update ($C$, \husky, 6.0.0, 7.0.4). We determine that client $C$ had integrated with \db in Section~\ref{\lblroot:data:project_selection}. Then, the \dbpr shown in Figure~\ref{\lblroot:fig:dbpr_example} is collected in Section~\ref{\lblroot:data:dependabot_pull_requests}, where we determine that \db has opened a \pr to update the \husky provider package. Next, in Section~\ref{\lblroot:data:compatibility_scores}, we collect the compatibility scores for \husky from the \db API. We end up collecting the compatibility scores not only for the 3-tuple update (\husky, \textit{6.0.0, 7.0.4}), but also all other 3-tuple combinations of (\husky, $V_C$, $V_T$) (e.g., (\husky, 6.0.1, 7.0.4), (\husky, 7.0.1, 7.0.4), (\husky, 6.0.0, 6.0.1), etc.). So, the compatibility scores for all of these 3-tuples we collected are included in the 3-tuple dataset. However, only the compatibility score for which \db has opened a \pr in $C$ (e.g., ($C$, \husky, 6.0.0, 7.0.4)) is included in the 4-tuple dataset.

		\section{Calculating a 90\% confidence interval for a compatibility score}
		\label{\lblroot:appendix:ci_calculation}

		We approach the task of calculating a confidence interval for a compatibility score using Bayesian inference, which is a statistical approach that aims at estimating a certain parameter (e.g., a mean or a proportion) from the population distribution, given the evidence provided by the observed (i.e., collected) data~\citep{hespanhol_understanding_2019}. In our case, we model the compatibility scores (i.e., the successful update ratio from the number of candidate updates) as a beta distribution, which defines random variables between 0 and 1, making it an ideal distribution choice for modelling the compatibility score~\citep{gupta_handbook_2004}. The beta distribution takes two parameters: $a$ and $b$. We start with a beta prior with $a$=1 and $b$=1 (which is a uniform prior), and our observed data of successful and failed counts for a dependency update. For a given true successful update ratio $p$ and $N$ candidate updates, the number of successful updates $S$ will look like a binomial random variable with parameters $p$ and $N$, where $N$ is the number of candidate updates and $p$ is unknown. This is because of the equivalence between successful update ratio and probability of a candidate update being a success or failure, out of $N$ possible candidate updates. So, with our Beta($a$=1, $b$=1) prior on $p$ and our observed successful updates $S$ $\sim$ Binomial($N$, $p$), then our posterior is also a beta distribution with $a=1+S$ and $b=1+N-S$.

		We use a normal approximation to calculate the standard deviation of the posterior beta distribution~\citep{gupta_beta_2011}. That is,
		\begin{equation}
		\label{\lblroot:eq:score_sigma}
		\begin{split}
			\sigma &= \sqrt{\frac{ab}{(a+b)^2(a+b+1)}} \\ 
			\text{where } a &= 1 + \textit{successful updates} \\
			b &= 1 + \textit{failed updates}
		\end{split}
		\end{equation}
		From there, we define a confidence interval precision as:
		\begin{equation}
		\label{\lblroot:eq:score_ci_precision}
		\textit{Precision}_{CI} = 1.65\sigma
		\end{equation}

		\noindent where $1.65$ is the critical (z) value (derived from the mathematics of the standard normal curve) to be used in confidence interval calculation associated with the 90\% confidence level~\citep{hazra_using_2017}.

		We define the bounds of the 90\% confidence interval as:
		\begin{equation}
		\begin{split}
		\label{\lblroot:eq:score_ci}
		\textit{CI} = [&\textit{max}(CS - \textit{Precision}_{CI}, 0), \\ 
		&\textit{min}(CS + \textit{Precision}_{CI}, 1)] \\
		\text{where } CS = \ \ &\textit{the existing compatibility score}
		\end{split}
		\end{equation}

\begin{figure}[ht]
			\centering
			\includegraphics[width=\linewidth]{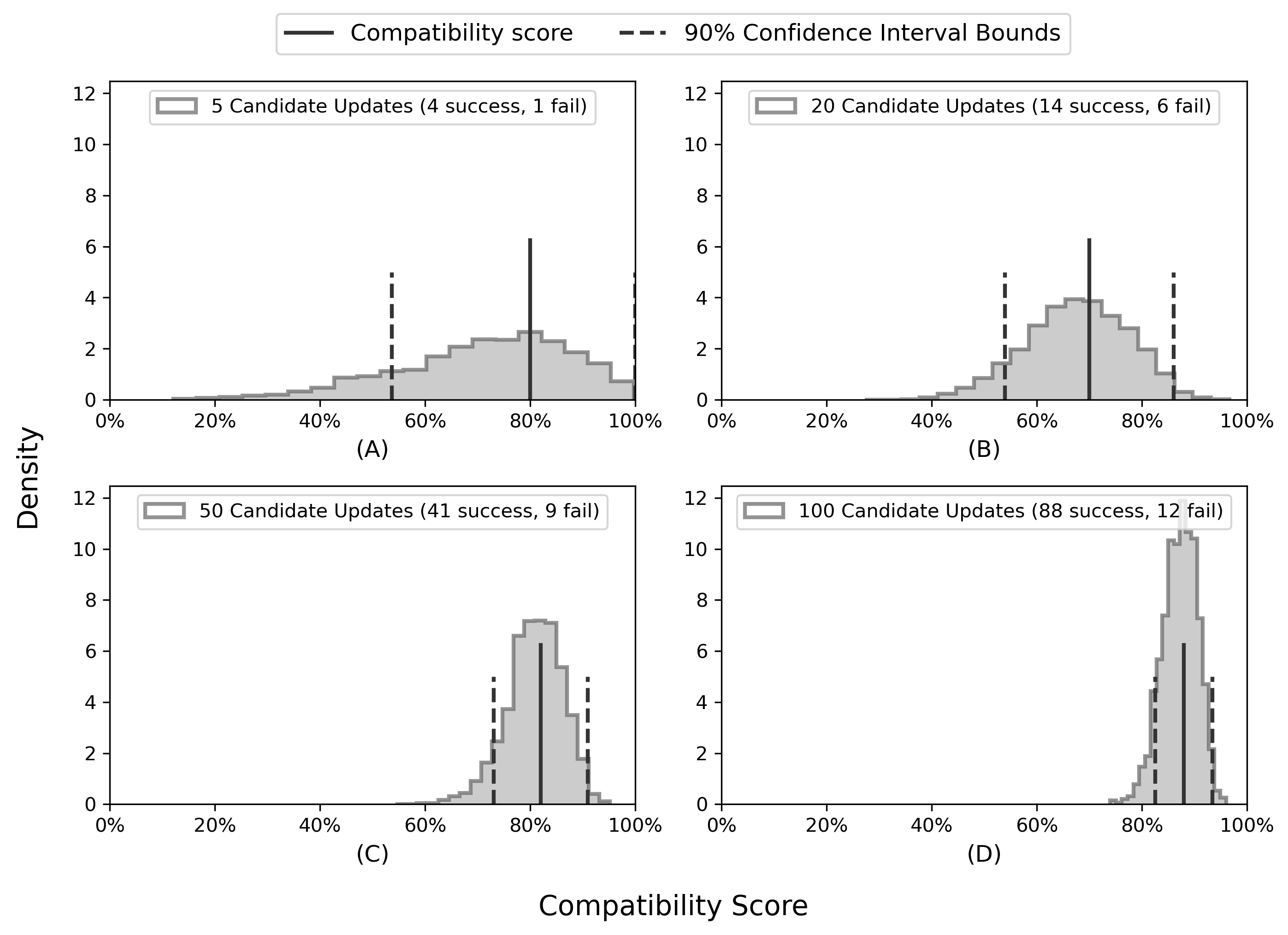}
			\caption{Examples of posterior distributions for particular success/failure update pairs with relatively low (A and B) and high (C and D) candidate updates. The vertical lines mark the compatibility score (solid) and the upper and lower bounds of the associated 90\% confidence interval (dashed).}
			\label{\lblroot:fig:sample_posterior_distributions_successful_update_ratio}
		\end{figure}

		Figure~\ref{\lblroot:fig:sample_posterior_distributions_successful_update_ratio} shows the posterior distributions resulting from the aforementioned process for particular success/failure update pairs. It can be seen that the distributions with a lower number of total candidate updates (i.e., A and B) have relatively wide distributions, expressing the uncertainly about what the true successful update ratio might be, whereas the distributions with a higher number of candidate updates (i.e., C and D) have tighter distributions. The solid vertical lines in Figure~\ref{\lblroot:fig:sample_posterior_distributions_successful_update_ratio} show the original compatibility score for each particular success/fail update pair, while the dashed vertical lines show the bounds of the associated 90\% confidence interval.

	\end{appendices}
    
\end{document}